\documentclass[12pt,a4paper]{article}
\usepackage{graphics}
\usepackage{amsmath}
\usepackage{amsfonts}
\input{psfig}
\newcommand{\newsec}[1]{\section{#1}}
 
\newcommand{\bpm}{\begin{pmatrix}}
\newcommand{\epm}{\end{pmatrix}}
\newcommand{\ba}{\begin{eqnarray}} \newcommand{\ea}{\end{eqnarray}}

\newcommand{\N}{{\cal N}}

\newcommand{\IZ}{{\mathbb Z}}

\newcommand{\half}{\frac{1}{2}}
\newcommand{\tr}{\operatorname{tr}}

\newcommand{\erf }{\operatorname{erf }}
\newcommand{\erfi}{\operatorname{erfi}}

\newcommand{\tpsi   }{{\tilde{\psi  }}}
\newcommand{\tphi   }{{\tilde{\phi  }}}
\newcommand{\ttheta }{{\tilde{\theta}}}

\newcommand{\hw}{\widehat{w}}
\newcommand{\hg}{\hat{g}}
\newcommand{\hepsilon}{\hat{\epsilon}}

\newcommand{\HNS }{      H^{\mbox {\tiny NS }}}

\newcommand{\HRR }{      H^{\mbox {\tiny RR }}}
\newcommand{\LQCD}{\Lambda_{\mbox {\tiny QCD}}}

\newcommand{\Ker  }{\operatorname{Ker  }}
\newcommand{\Image}{\operatorname{Image}}
\newcommand{\Span }{\operatorname{Span }}

\newcommand{\tmin}{{\mbox {\tiny min}}}

\begin{document}

% title page 
\begin{titlepage}
%\rightline{}
\rightline{\tt hep-th/0201224}
\rightline{TAUP-2695-02}
\rightline{WIS/05/02-JAN-DPP}
%\rightline{\today}
\vskip 0.5cm
\centerline{{\Large \bf }} 
\vspace{6pt}
\centerline{{\Large \bf Stable Non--Supersymmetric Supergravity Solutions }}
\vspace{6pt}
\centerline{{\Large \bf from Deformations of the Maldacena--Nu\~nez Background }}
\vskip 1cm
\centerline{Ofer Aharony$^{a,}$\footnote{E-mail :
{\tt Ofer.Aharony@weizmann.ac.il.}
Incumbent of the Joseph and Celia Reskin
career development chair.}, Ehud Schreiber$^{b,}$\footnote{E-mail : 
{\tt schreib@post.tau.ac.il.}} and Jacob Sonnenschein$^{b,c,}$\footnote{
E-mail : {\tt cobi@post.tau.ac.il.}}}
\vskip 1cm
\begin{center}
\em $^a$ Department of Particle Physics,
\\The Weizmann Institute of Science, Rehovot 76100, Israel
\end{center}
\begin{center}
\em $^b$ School of Physics and Astronomy,
\\Beverly and Raymond Sackler Faculty of Exact Sciences,
\\Tel Aviv University, Ramat Aviv, 69978, Israel
\end{center}
\begin{center}
\em $^c$ School of Natural Sciences\footnote{Address for academic year
  2001--2002.},
Institute for Advanced Studies,
\\Einstein Drive, Princeton, New Jersey, 08540, USA
\end{center}
\vskip 0.7cm 
We study a deformation of the type IIB Maldacena--Nu\~nez
background which arises as the near--horizon limit of NS5 branes
wrapped on a two--cycle. This background is dual to a ``little string
theory" compactified on a two--sphere, a theory which at low energies
includes four--dimensional ${\cal N}=1$ super Yang--Mills theory.  The
deformation we study corresponds to a mass term for some of the scalar
fields in this theory, and it breaks supersymmetry completely. In the
language of seven--dimensional $SO(4)$ gauged supergravity the
deformation involves (at leading order) giving a VEV, depending only
on the radial coordinate, to a particular scalar
field.  We explicitly construct the corresponding solution
at leading order in the deformation, both in seven--dimensional and in
ten--dimensional supergravity, and we verify that it
completely breaks supersymmetry. Since the original background had a
mass gap and we are performing a small deformation,
the deformed background is guaranteed to be stable even though it is
not supersymmetric.

\end{titlepage}

\newsec{Introduction}

\subsection{The Quest for a Non--Supersymmetric Stable Background}

The principle of holography states that any theory of gravity in $d$
dimensions is equivalent to a non--gravitational theory in $d-1$
dimensions. This principle is completely independent of supersymmetry,
but its first explicit realizations in the AdS/CFT correspondence
\cite{Maldacena:1997re,Gubser:1998bc,Witten:1998qj,Aharony:1999ti} 
were all supersymmetric. Attempts to find non--supersymmetric
holographic dual pairs starting from the gravitational side, for
instance by looking at non--supersymmetric solutions involving an AdS
space, encountered the same problems as attempts to find stable
non--supersymmetric string theory backgrounds (see e.g. \cite{Berkooz:1998qp}
for a discussion). One problem is that loop effects generally generate 
a potential for the dilaton and any other moduli scalars, 
which tends to destabilize the vacuum. 
Another problem is that such backgrounds generically involve
tachyonic fields which do not satisfy the Breitenlohner--Freedman (BF)
bound, and these also tend to destabilize such spaces. Both types of
problems generally show up in the particular example of
non--supersymmetric orbifolds \cite{Kachru:1998ys}.

There are various ways to get around these problems and find
non--supersymmetric holographic dual pairs. One possibility is to find
non--supersymmetric backgrounds which do not have uncharged moduli or tachyons
below the BF bound. Such backgrounds do not occur as perturbative
string theory solutions, since in string theory the dilaton is always
a modulus, but there are many such backgrounds in M theory; for
instance, supersymmetric M theory backgrounds of the form $AdS_4\times M^7$
(where $M^7$ is a compact space which is not $S^7$) always have
non--supersymmetric twins related to them by
``skew--whiffing'' \cite{Duff:nu}, and other stable examples also exist. The
main problem with such M--theoretic examples is that the corresponding
field theories are poorly understood, and it is not known how to get
theories resembling QCD in this way.

Another possibility is to start from a supersymmetric background which
is dual to a known field theory, and to deform the field theory in a
way which breaks supersymmetry; for instance, one can start from a
superconformal theory and deform it by a relevant deformation. The
simplest example involves starting from ${\cal N}=4$ super Yang--Mills
(SYM) theory and adding a mass term for the scalars. The problem with
this example (and with other examples of this type) is that giving an
equal mass to all the scalars corresponds to deforming by a non--chiral
operator
which has a large anomalous dimension and is not relevant in the
supergravity regime, while deformations which are relevant always
involve giving a negative mass--squared to at least one scalar, so
they do not lead to a stable non--supersymmetric vacuum. Generally,
when one is deforming a theory with a moduli space, one must make
sure that the deformation does not generate a potential on the moduli
space which would destabilize the vacuum. Since the theories which
have supergravity duals are usually strongly coupled, it is difficult
to analyze this question. However, several examples are known of
deformations which do lead to stable non--supersymmetric backgrounds,
starting from \cite{Polchinski:2000uf} where the ${\cal N}=4$ theory
was deformed by fermion mass terms. For some values of the masses this
deformation breaks
supersymmetry and still leads to a stable vacuum
\cite{Polchinski:2000uf,Zamora:2000ha,Kinar:2000wh}. This example (and
similar ones) involves fivebranes, so it cannot be fully described by
supergravity. Another attempt to break supersymmetry using exactly
marginal deformations of two dimensional CFTs was described in
\cite{Aharony:2001dp}; in this case the deformation leads to a
non--local theory on the worldsheet, so again one does not get a
duality between a standard perturbative string theory and a known field
theory. Examples of meta--stable non--supersymmetric backgrounds were
recently discussed in \cite{Kachru:2001gs}. Another way to break
supersymmetry is by adding a finite temperature in the field theory,
following \cite{Witten:1998zw}, but here we focus only on
Lorentz--invariant configurations.

In this paper we construct the first example of a stable
non--supersymmetric supergravity background which is holographically
dual to a field theory including four--dimensional Yang--Mills (YM)
theory\footnote{More precisely, this is the first example (as far as
we know) which is the low--energy limit of a string theory which can
be chosen to be weakly coupled everywhere; an example involving strong
coupling, coming from compactified D4 branes, 
was described in \cite{Witten:1998zw}.}.  The field theory in
this case is a deformation of ``little string theory'' (see
\cite{Aharony:1999ks} for a review) compactified on a two--sphere,
whose holographic dual was found in \cite{MalNun}. This theory arises
for instance from a decoupling limit of $N$ NS5 branes wrapped on a
two--cycle in a Calabi--Yau manifold.  At
low energies (in the case of type IIB NS5 branes) this theory includes
four--dimensional YM theory with gauge group $SU(N)$, though in the
supergravity approximation these modes are inseparable from other
modes coming from the compactification \cite{MalNun}. As found in
\cite{MalNun}, this background has a mass gap, so we are guaranteed
that it will be stable under small deformations, even if they break
supersymmetry. In this paper we will describe in detail a particular
deformation of this theory, corresponding to a six--dimensional mass
term for scalar fields.  The particular example we analyze is not in
the same universality class as pure YM theory because it includes a
massless adjoint fermion with a $U(1)$ R--symmetry (classically in the
UV) which protects it
from acquiring a mass; however, it should be possible also to
construct generalizations of our example which could be in the same
universality class as pure YM. In our case, as in other holographic
constructions, going to a limit corresponding to a string theory for
pure YM theory or QCD (which is our eventual goal), without any
additional fields, requires going beyond the supergravity
approximation and performing a full string theory analysis, which is
beyond our current capabilities.

\subsection{The Maldacena--Nu\~nez Supersymmetric Background}

A stack of $N$ flat NS5 branes in type IIB string theory gives rise to
a linear dilaton background, which is dual, in an appropriate
decoupling limit, to a ``little string theory'' \cite{ABKS}.  At low
energies this theory includes a six--dimensional ${\cal N}=(1,1)$ SYM
theory with $SU(N)$ gauge group, which includes four adjoint scalars
$\Phi_a$.  Maldacena and Nu\~nez \cite{MalNun} analyzed the
supergravity solution corresponding to wrapping the fivebranes on a
two--sphere $S^2$, with the remaining four directions spanning a
four--dimensional Minkowski space. By twisting the normal bundle
appropriately (this is automatic when the fivebranes wrap a two--cycle
in a Calabi--Yau manifold) one fourth of the original supersymmetry is
maintained, corresponding to four--dimensional ${\cal N}=1$ supersymmetry, and
the scalars $\Phi_a$ all become massive\footnote{Analogous solutions in
which the compactification maintains half of the supersymmetry were 
analyzed in \cite{GKMW,BiCoZa}.}.  At low energies this theory
includes the four--dimensional ${\cal N}=1$ pure SYM theory. The two--sphere
which the theory is compactified on appears also in the dual
supergravity background, and its radius decreases as we go from the UV
(large radial coordinate) to the IR (small radial
coordinate). 
%However, the radius remains large also in the IR,
%reflecting the fact that Kaluza--Klein modes related to the reduction
%on the sphere are not decoupled from the $d=4$ SYM fields.

% (the mass
%gap of the SYM field theory is of the same order of magnitude as the
%inverse radius of the two--sphere).

%The directions transverse to the branes are the radial distance $\rho$
%and a three--sphere $S^3$. 
%In the ultraviolet (UV) region $\rho \rightarrow \infty$,
%the radius of the two--sphere is large,
%the field theory is six--dimensional and is similar to that of the 
%flat branes case.
%In the infrared (IR) region 
%$\rho \rightarrow 0$, the radius of the two--sphere is smaller, and the theory
%gives rise to four--dimensional ${\cal N} = 1$ SYM theory without any
%massless matter. Still, there are also Kaluza--Klein modes apart from
%the SYM fields.
%The change of the two--sphere radius reflects the
%running of the coupling constant of the four--dimensional theory. 

All the supergravity fields which are non--zero in the solution are
contained in the ten--dimensional type I supergravity sector of the
type IIB theory. For fivebranes in flat space, 
the isometries of the transverse three--sphere give rise to an
$SO(4) \equiv SU(2)_L \times SU(2)_R$ symmetry which is the
R--symmetry of the corresponding theory. In the theory of the
wrapped fivebranes, the remaining R--symmetry is just a $U(1)$
subgroup of $SU(2)_L$.  Reducing type I supergravity on the
three--sphere gives rise to a seven--dimensional $SO(4)$ gauged
supergravity \cite{FreSch}, which is a consistent truncation of the
full theory. The four--dimensional Minkowski space is merely a spectator
in the solution, so the relevant gauged supergravity can be thought of
as three--dimensional, or four--dimensional if we keep the time
direction.

In the seven dimensional supergravity description,
the twisting of the normal bundle is achieved by turning on one of the
$SU(2)_L$ gauge
fields, which is taken to be $A_L^3$. For the solution of \cite{MalNun}
it is sufficient to work with a supergravity which is further truncated,
and includes only the $SU(2)_L$ gauge fields, since no fields charged
under $SU(2)_R$ participate in the solution.
The naive solution in which one turns on only $A_L^3$
is analogous to an $SU(2)$ Dirac
monopole, and is unphysically singular. However, one can find a smooth
solution with the same asymptotic behaviour
by turning on also the $A_L^1, A_L^2$ gauge fields. This was carried out in
the four--dimensional $SU(2)$ gauged supergravity context by 
Chamseddine and Volkov \cite{ChaVol1, ChaVol2}. 
Maldacena and Nu\~nez translated that solution to
the context of seven--dimensional supergravity, and raised the
solution to ten dimensions using the results of \cite{CvLuPo, ChaSab},
to get a smooth solution describing the fivebranes compactified on $S^2$.

The singular solution exhibits a classical $U(1)$ R--symmetry which is broken
by anomalies to $\IZ_{2 N}$. The instantons responsible for this are
given, roughly, by fundamental string worldsheets wrapped on the
two--sphere \cite{MalNun}. 
The smoothed solution further breaks the symmetry spontaneously to
$\IZ_2$, as expected from the field theory point of view. Maldacena and
Nu\~nez also S--dualized the solution to represent wrapped D5 branes,
and found the behaviour of the string tension, glueball masses and
domain wall tensions of the SYM theory. As usual
in such cases, when the supergravity approximation is valid, the typical
mass of the lightest glueballs, $\LQCD$, is of the same order as the masses of
the Kaluza--Klein modes on the two--sphere, and the SYM theory is not
really decoupled from those additional degrees of freedom.

\subsection{The Supersymmetry Breaking Perturbation}

In this paper we explore the supergravity solution which is dual to
explicitly breaking the supersymmetry in the ``little string theory''
(LST). At low energies, the ${\cal N}=(1,1)$ LST coming from $N$ type
IIB NS5 branes reduces to the six dimensional ${\cal N}=(1,1)$ $SU(N)$
SYM theory, which is the dimensional reduction of ten--dimensional SYM
to six dimensions. This theory includes four adjoint scalar fields
$\Phi_a$ in the $\bf(2,2)$ representation of the global $SO(4)$
R--symmetry. The simplest chiral operator in the theory, as in other
maximally supersymmetric SYM theories, is ${\cal O} = X^{ab} \tr
\Phi_a \Phi_b$ where $X$ is a traceless symmetric $SO(4)$ matrix; this
operator is in the $\bf(3,3)$ representation of $SO(4)$. In appendix A
we define a basis $X_{lr}$ for this representation, where the indices
$l,r$ are in the adjoint of $SU(2)_{L,R}$, respectively. We wrote the
operator $\cal O$ as an operator of the low--energy SYM theory, but it
is actually a chiral operator in the full LST
\cite{ABKS}, which reduces to this form at low energies. Our
deformation will involve adding a term of the form $\epsilon {\cal O}$
to the Lagrangian of the six--dimensional theory, and we will find the
dual background at leading order in perturbation theory in $\epsilon$
(the dimensions will be set by the string tension, which is also the
inverse Yang--Mills coupling in the six--dimensional theory, or by the
radius of the two--sphere on which the theory is compactified).

In the six--dimensional theory we can always diagonalize the matrix $X$ 
by a global $SO(4)$ transformation. A deformation by $\cal O$ is thus
determined by
three invariants, or alternatively by four eigenvalues whose sum is zero. 
In the compactified theory the $SU(2)_L$ is broken to a $U(1)$ which
is the R--symmetry group (in our conventions, this $U(1)$ corresponds 
to the adjoint index $l = 3$ of $SU(2)_L$). 
Clearly, there is a difference between choosing 
the deformation to be charged or uncharged under this $U(1)$. 
In this paper we will analyze the uncharged case. In this case the matrix
$X$ appearing in the deformation is of the form $v^r X_{3r}$  
(in the conventions of appendix A) for some arbitrary vector $v^r$, 
and its eigenvalues are proportional to $\{-1,-1,+1,+1\}$.
Without loss of generality, we may
take $X \equiv X_{33}$ and the corresponding operator is 
${\cal O} = \tr \left( \Phi_1^2 - \Phi_2^2 - \Phi_3^2 + \Phi_4^2 \right)$. 
As our deformation will preserve the classical
$U(1)$ R--symmetry in the UV, its breaking will remain as in
\cite{MalNun}. In particular, this symmetry forbids (before it is
spontaneously
broken) the generation of a mass term for the gluino in the
four--dimensional ${\cal N}=1$ SYM multiplet, so even after the deformation we
will have (classically) a massless four--dimensional adjoint fermion,
despite the absence of supersymmetry.

From the point of view of the six--dimensional theory we are giving
some of the scalars in the SYM theory a negative mass squared, making
them tachyonic. Thus, if we try to perform this SUSY--breaking
deformation directly in the six--dimensional theory, it will
destabilize the vacuum and the theory will run to large values of
$\Phi_1$ and $\Phi_4$ (if $\epsilon > 0$). However, when we compactify
on $S^2$ to four dimensions, all the modes of the fields $\Phi_a$ become
massive, with a mass of at least the order of the inverse
compactification scale. Thus, as long as $\epsilon$ is small enough,
the deformation in the compactified theory does not destabilize the
vacuum, but just changes the masses of the already massive
fields. Another way to see that the resulting theory must be stable (at
least for small values of $|\epsilon|$) is to note that the theory
before the deformation had a mass gap, so it cannot be destabilized by
any small deformation. 

Since the fermion masses get no contribution at leading order in
$\epsilon$, supersymmetry is explicitly broken by the deformation.
The fields in the dual background which are dual to the operator $\cal
O$ were described in \cite{ABKS} for the uncompactified six
dimensional theory; they involve a squashing of the metric on the
3--sphere. In the seven--dimensional supergravity theory they are nine
scalar fields $c_{lr}$.
% which must be included in the Lagrangian. 
Since the operator ${\cal O}$ is charged under
both $SU(2)_L$ and $SU(2)_R$, we can no longer work (as in
\cite{MalNun}) with the truncation of the seven--dimensional supergravity
to $SU(2)_L$, but have to deal with the full
$SO(4)$ gauged supergravity. Since supersymmetry is broken, we can't
use the BPS first order differential equations to find the new
solution, and we must deal with the second order equations of motion.

The organization of this paper is as follows. 
In the next section we translate the 
$SO(4)$ gauged supergravity Lagrangian of \cite{CvLuPo} to the 
seven--dimensional $SU(2)_L \times SU(2)_R$ language, 
including the nine scalars and working to quadratic order in $\epsilon$. 
Section 3 describes the singular four--dimensional monopole
solution and the smooth monopole solution of Chamseddine and Volkov, 
and studies their symmetries. 
In section 4 we derive the order $\epsilon$ equations of motion 
for the perturbation around the Chamseddine--Volkov solution, and
obtain the constraints imposed by respecting the symmetries.
In section 5 we specialize to the most symmetric case, where the
scalar depends only upon the radial direction $\rho$, and find the
solutions for the singular and the non--singular cases. 
In section 6 the solution is raised to ten dimensions according to
\cite{CvLuPo}. 
In section 7 we S--dualize the solution and look at the string tension. 
%In section 8 we show how the classical $U(1)$ R--symmetry, which is
%maintained in the perturbed solution, is anomalously broken.
In section 8 we explicitly show that our ten--dimensional supergravity background breaks supersymmetry. 
We conclude with a summary and discussion. 
Some matrix conventions and details of the calculations are relegated
to three appendices.

\newsec{The Supergravity Lagrangian}
\label{sec:lagrangian}

The authors of \cite{CvLuPo} study consistent reductions
of a $D$ dimensional theory, containing the metric, a dilaton and a
Kalb--Ramond field, down to $D-3$ dimensions, by the Kaluza--Klein
mechanism of compactification on an $S^3$. We are, of course,
interested in the case of $D = 10$ supergravity, giving rise to 
seven--dimensional, $SO(4)$ gauged supergravity. 
The massless fields in this
theory are the metric, a scalar $Y$, the $SO(4)$ gauge fields 
$A^{ab}_{(1)}$, a symmetric unimodular matrix $\tilde{T}_{ab}$ 
(that is $\tilde{T}_{ab} = \tilde{T}_{ba}$ and $\det \tilde{T} = 1$),
and also a two--form potential $A_{(2)}$. 
We will denote the $SO(4)$ gauge coupling constant by $\hg$.

The indices $a,b,c,d,\ldots = 1,2,3,4$ are $SO(4)$ vector
indices. $A_{(1)}$ is in the adjoint representation, ${\bf 6}$, of
$SO(4)$, 
\begin{equation}
\label{Aij}
A^{ab}_{(1)} = - A^{ba}_{(1)}, 
\end{equation}
while $\tilde{T}$ is in the representation ${\bf 9}$.
The $SO(4)$ decomposes to $SU(2)_L \times SU(2)_R$ where the left
(right) subgroups have (anti) self--dual gauge potentials:
\begin{equation}
A^{ab}_{(1)} = \pm \half \epsilon_{abcd} A^{cd}_{(1)}.
\end{equation} 
We will denote the left and right $SU(2)$ gauge potentials as 
$A_L^l$ and $A_R^r$ respectively, 
where $l,m,n,\ldots = 1,2,3$ are $SU(2)_L$ adjoint
indices, or equivalently $SO(3)$ vector indices, and similarly
$r,s,t,\ldots = 1,2,3$ for $SU(2)_R$.

Under this decomposition, the gauge potential representation obviously
decomposes as ${\bf 6} = ({\bf 3} , {\bf 1}) + ({\bf 1} , {\bf 3})$,
while ${\bf 9} = ({\bf 3} , {\bf 3})$. We choose a certain
embedding of the $SU(2)_L$ and $SU(2)_R$ generators, $\alpha_L^l$
and $\alpha_R^r$ respectively, in the $SO(4)$ adjoint
representation. We also choose a specific representation $X_{lr}$ of
the tangent space of the $\tilde{T}$ at the origin $\tilde{T} = \delta_{ab}$ 
(for each $l = 1,2,3$ and $r = 1,2,3$, $X_{lr}$ is a $4 \times 4$
traceless matrix $(X_{lr})_{ab}$). The details are described in appendix A. 
We define the fields $c_{lr}$ by the parameterization 
\begin{equation}
\label{tildeT}
\tilde{T} = \exp Q,
\end{equation}
where 
\begin{equation}
\label{trQ}
Q = c_{lr} X_{lr},\ \tr Q = 0.
\end{equation} 
We will also denote by
$c$ the corresponding $3 \times 3$ matrix.

Let us define the field strengths 
\begin{equation}
F_{(2)}^{ab} = d A_{(1)}^{ab} + \hg A_{(1)}^{ac} \wedge A_{(1)}^{cb}
\end{equation}
and
\begin{equation}
\label{F3}
F_{(3)} = d A_{(2)} +
\frac{1}{8} \epsilon_{abcd} (F_{(2)}^{ab} \wedge A_{(1)}^{cd} - 
        \frac{1}{3} \hg A_{(1)}^{ab} \wedge A_{(1)}^{ce} \wedge A_{(1)}^{ed}),
\end{equation}
as well as the covariant derivative of $\tilde{T}_{ab}$,
\begin{equation}
\label{DT}
{\cal D} \tilde{T}_{ab} = 
d \tilde{T}_{ab} + \hg (A_{(1)}^{ac} \tilde{T}_{cb} + 
                        A_{(1)}^{bc} \tilde{T}_{ac}   ), 
\end{equation}
compatible with its $SO(4)$ index structure.

The Einstein frame Lagrangian of the $(D-3)$--dimensional
 massless theory was found to be
\cite{CvLuPo}, in the language of differential forms, 
\begin{eqnarray}
{\cal L}_{D-3} & = & R \star {1} 
    - \frac{D-5}{16} Y^{-2} \star d Y \wedge d Y 
    - \frac{1}{4} \tilde{T}_{ab}^{-1} \star {\cal D} \tilde{T}_{bc} \wedge
                  \tilde{T}_{cd}^{-1}       {\cal D} \tilde{T}_{da}
                  \nonumber \\
& & - \frac{1}{2} Y^{-1} \star F_{(3)} \wedge F_{(3)}
    - \frac{1}{4} Y^{-1/2} \tilde{T}_{ac}^{-1} \tilde{T}_{bd}^{-1} 
                           \star F_{(2)}^{ab} \wedge F_{(2)}^{cd} 
    - V \star {1}, 
\end{eqnarray}
where the potential $V$ is given by
\begin{equation}
V = \frac{1}{2} \hg^2 Y^{1/2} \left( 2 \tilde{T}_{ab} \tilde{T}_{ab} 
                                     - (\tilde{T}_{aa})^2           \right).
\end{equation}

The first term in the Lagrangian is the Einstein--Hilbert term. The
second is the kinetic energy of the scalar $Y$, the third is that of
the scalars in the ${\bf 9}$ representation, while the fourth is that
of $A_{(2)}$. The fifth term is the Maxwell term.

Let us look at the $A_{(2)}$ kinetic term. If we truncate the gauge
group to the diagonal $SU(2)_D \subset SU(2)_L \times SU(2)_R$, 
as in \cite{GKMW, BiCoZa}, the scope
of the indices $a,b,c,d,\ldots$ is reduced to (say) $1,2,3$. Thus
$\epsilon_{abcd}$ vanishes in (\ref{F3}), and we are left with 
$F_{(3)} = d A_{(2)}$. The Lagrangian is therefore
quadratic in $A_{(2)}$, and $A_{(2)}$ can be decoupled from
the rest of the fields. Naively, for the full $SO(4)$ theory
this decoupling is not possible. 
However, in the case $D = 10$ this theory can also be written \cite{ChaSab} 
in an equivalent form 
\cite{TowNie,Quevedo:fv}, using a field $A_{(3)}$ dual to $A_{(2)}$.  
The Lagrangian is given in equation (19) of \cite{ChaSab}. 
We will be interested in the ansatz in which the seven--dimensional
space is a product of a four--dimensional Minkowski space and a 
three--dimensional space, and where the four--dimensional Poincar\'e
invariance is maintained. The latter form of the Lagrangian makes it
easy to see that in this case, the $A_{(3)}$ field has no sources from
the other fields. Therefore, its action is quadratic and it can be
decoupled. Thus, we will henceforth ignore this extra field. This
has also been done in the analysis of \cite{MalNun}, where the gauge
group is truncated to $SU(2)_L$.

We will now write the scalar Lagrangian $L_7$,
\begin{equation}
{\cal L}_7 = \sqrt{\det g_{\mu \nu}} \; L_7 \; d x^0 \wedge d x^1 \wedge
\cdots \wedge d x^6,
\end{equation} 
using the standard definitions
\begin{eqnarray}
A_{(1)}^{ab} & = & A_\mu^{ab} d x^\mu,  \\
F_{\mu \nu}^{ab} & = & \partial_\mu A_\nu^{ab} - \partial_\nu A_\mu^{ab} 
                  + \hg (A_\mu^{ac} A_\nu^{cb} - A_\nu^{ac} A_\mu^{cb}),  \\
{\cal D}_\mu \tilde{T}_{ab} & = &
\partial_\mu \tilde{T}_{ab} + \hg (A_\mu^{ac} \tilde{T}_{cb} + 
                                   A_\mu^{bc} \tilde{T}_{ac}   ). 
\end{eqnarray}
We get
\begin{eqnarray}
L_7 & = & R - \frac{5}{16} Y^{-2} \partial_\mu Y \partial^\mu Y 
          - \frac{1}{4} \tilde{T}_{ab}^{-1} {\cal D}_\mu \tilde{T}_{bc} 
                        \tilde{T}_{cd}^{-1} {\cal D}^\mu \tilde{T}_{da}
                  \nonumber \\
    &  & - \frac{1}{8} Y^{-1/2} \tilde{T}_{ac}^{-1} \tilde{T}_{bd}^{-1} 
                       F_{\mu \nu}^{ab} F^{cd \mu \nu}            
         - V.
\label{L7preSO4}
\end{eqnarray}
Notice the numerical coefficient $\frac{1}{8}$ in the Maxwell
term. This is, however, the canonical normalization, in light of (\ref{Aij}). 
When truncating to $SU(2)_D$, by limiting the scope of the indices
$a, b,\ldots$, this normalization is kept. However, when decomposing
$SO(4)$ as $SU(2)_L \times SU(2)_R$, or when truncating to $SU(2)_L$,
the normalization is non--standard. 
Thus the gauge coupling constant $g$ of the $SU(2)$ groups will differ
from the $SO(4)$ gauge coupling constant $\hg$.

We define the $SU(2)_L$ and $SU(2)_R$ gauge fields, 
$A_L^l$ and $A_R^r$ respectively, by
%\begin{equation}
%-i A^{ij} E^{ij} = Sqrt{2} (A_L^l \alpha_L^l + A_R^r \alpha_R^r) 
%\end{equation}
%Where $(E^{ij})$ is the $4 \times 4$ matrix having entry $1$ in the
%$i$-th row at the $j$-th column, and zeroes otherwise, 
%$(E^{ij})_{kl} = \delta_{ik} \delta{jl}$.
\begin{equation}
-i A = \sqrt{2} (A_L^l \alpha_L^l + A_R^r \alpha_R^r), 
\end{equation}
where $A$ is viewed as a $4 \times 4$ matrix in the indices $a,b$,
besides the spatial index $\mu$. 
Viewing likewise $F$ as a $4 \times 4$ matrix, we get
\begin{equation}
-i F = \sqrt{2} (F_L^l \alpha_L^l + F_R^r \alpha_R^r), 
\end{equation}
with
\begin{eqnarray}
F_{L \mu \nu}^l & = &  \partial_\mu A_{L \nu}^l - \partial_\nu A_{L \mu}^l 
                     - \sqrt{2} \hg \epsilon_{lmn} A_{L \mu}^m A_{L \nu}^n, \\
F_{R \mu \nu}^r & = &  \partial_\mu A_{R \nu}^r - \partial_\nu A_{R \mu}^r 
                     - \sqrt{2} \hg \epsilon_{rst} A_{R \mu}^s A_{R \nu}^t. 
\end{eqnarray}

Apart from reverting to the $SU(2)_L \times SU(2)_R$ language, we also wish  
to expand the Lagrangian near the origin $\tilde{T}_{ab} = \delta_{ab}$,
namely, to expand in powers of $c$. The field $c$ is dual to the
operator ${\cal O}$ in LST, and when we deform the LST Lagrangian by 
$\epsilon {\cal O}$, $c$ will be proportional to $\epsilon$, which we
take to be small. In order to solve the equations of motion up to
first order in $\epsilon$, it suffices to keep terms in the Lagrangian
up to second order in $c$.
In appendix B we calculate the various terms of the Lagrangian 
(\ref{L7preSO4}) to second order in the fields
$c_{lr}$ and their derivatives.
We will also substitute 
\begin{equation}
Y = e^y,
\end{equation}
where $y$ is proportional to the dilaton.

The final form of the Lagrangian,
gathering all the terms and defining
\begin{equation}
g \equiv - \sqrt{2} \hg 
\end{equation}
in order to get the standard coupling, is
\begin{eqnarray}
\label{L7SO4}
L_7 & = & R - \frac{5}{16} \partial_\mu y \partial^\mu y 
            - {\cal D}_\mu c_{lr} {\cal D}^\mu c_{lr}  \nonumber \\
    &   &   -\frac{1}{4} e^{-y/2} (F_L^l F_L^l + F_R^r F_R^r + 
                                   4 F_L^l F_R^r c_{lr} 
                                   + 2 (F_L^l F_L^m c_{lr} c_{mr}  + 
                                        F_R^r F_R^s c_{lr} c_{ls})   )
                  \nonumber \\
    &   &   + 2 g^2 e^{y/2} + O(c^3),
\end{eqnarray}
where
\begin{eqnarray}
F_{L \mu \nu}^l & = &   \partial_\mu A_{L \nu}^l - \partial_\nu A_{L \mu}^l 
                      + g \epsilon_{lmn} A_{L \mu}^m A_{L \nu}^n,    \\
F_{R \mu \nu}^r & = &   \partial_\mu A_{R \nu}^r - \partial_\nu A_{R \mu}^r 
                      + g \epsilon_{rst} A_{R \mu}^s A_{R \nu}^t,    \\
{\cal D}_\mu c_{lr} & = & \partial_\mu c_{lr}  
    + g (\epsilon_{lmn} A_{L \mu}^m c_{nr} + 
         \epsilon_{rst} A_{R \mu}^s c_{lt}   ),   
\end{eqnarray}
and where, for example, $F_L^l F_L^l \equiv F_L^{l \mu \nu} F_{L \mu \nu}^l$.

\newsec{The Chamseddine--Volkov Solution}
The Lagrangian (\ref{L7SO4}) can be truncated to an $SU(2)_L$ gauged
supergravity theory. Indeed, it is an $SU(2)_R$ invariant, so all the
fields charged under $SU(2)_R$ appear at least quadratically in all the
terms, and can be decoupled. The resulting Lagrangian after decoupling
the fields $A_R^r$ and $c_{lr}$ is
\begin{equation}
\label{L7SU2}
L_7 = R - \frac{5}{16} \partial_\mu y \partial^\mu y 
      -\frac{1}{4} e^{-y/2} F_L^l F_L^l
      + 2 g^2 e^{y/2}. 
\end{equation}
We will look for warped geometry solutions having a four--dimensional
Minkowski space factor (with coordinates $x_\mu$). 
Those solutions are therefore essentially
three--dimensional, and were investigated in the context of 
four--dimensional supergravity \cite{ChaVol1,ChaVol2}. 
Moreover, the remaining three--dimensional
geometry will have an $SO(3)$ symmetry, and will be a warped product of
a two--sphere (with coordinates $\theta, \phi$) and a half line (the radial
coordinate $\rho$). 
The general form of the geometry in the string frame, modulu the
reparameterization invariance of $\rho$, corresponding to
$N$ wrapped NS5 branes \cite{MalNun} is 
\begin{equation}
d s_{7,\mbox{st}}^2 = dx_\mu dx^\mu + N
  \left( d \rho^2 + e^{2 h(\rho)} (d \theta^2 + \sin^2 \theta d \phi^2) \right).
\end{equation}
We will work in the Einstein frame, whose metric is given by 
\begin{equation}
d s_7^2 = e^{-y(\rho)/2} \, d s_{7,\mbox{st}}^2.
\end{equation}
By looking at the radius of the transverse $S^3$ we find that
$N$ is related to the coupling constant $g$ of the gauged supergravity by
\begin{equation}
\label{Nofg}
N = \frac{2}{g^2}. 
\end{equation}

There is a solution involving only the gauge field $A_L^3$ (the
Lagrangian can be further truncated to the theory containing
only that $U(1) \subset SU(2)_L$). It reads
\begin{eqnarray}
\label{e2hrhosing}
e^{2 h(\rho)} & = & \rho, \\
\label{yrhosing}
y(\rho)       & = & \frac{8}{5} 
   \left( \varphi_0 - \rho + \frac{1}{4} \log \rho \right), \\
\label{AL3sing}
A_L^3         & = & \frac{1}{g} \cos \theta d \phi,
\end{eqnarray}
(that is, $A_{L \phi}^3 = (\cos \theta) / g$ and all its other
components vanish). 
This solution, however, is singular at $\rho = 0$, and thus
it is unphysical. 

Let us examine the symmetries of this solution. The field strength is
\begin{equation}
\label{FL3singular}
F_{L}^3 = -\frac{1}{g} \omega,
\end{equation}
where
\begin{equation}
\label{volumeform}
\omega = \sin \theta \, d \theta \wedge d \phi
\end{equation}
is the $SO(3)$ invariant volume form of $S^2$. This enables the whole 
solution to be $SO(3)$ invariant, and in particular the scalars $y$ and $h$
to be radial functions only. 
%At the level of the gauge vector, 
The symmetry under an infinitesimal rotation by $\hepsilon$ around the $z$ 
axis,
\begin{eqnarray}
\delta_3 \theta & = & 0,        \\
\delta_3 \phi   & = & \hepsilon,
\end{eqnarray}
is obvious, as $A_L^3$ depends on $\phi$ only through $d \phi$ which is 
invariant. Let us now look at a rotation around the $y$ axis,
\begin{eqnarray}
\delta_2 \theta & = &   \hepsilon             \cos \phi, \\
\delta_2 \phi   & = & - \hepsilon \cot \theta \sin \phi.
\end{eqnarray}
This time $A_L^3$, given by (\ref{AL3sing}), is not invariant, 
but the change is a gauge transformation,
\begin{equation}
\delta_2 A_L^3 = \frac{1}{g} \, d \Lambda,
\end{equation}
with
\begin{equation}
\label{Lambda}
\Lambda = -\hepsilon \, \frac{\sin \phi}{\sin \theta}.
\end{equation}
Of course, $\delta_2$ and $\delta_3$ generate the whole $SO(3)$ symmetry.

The singular solution can be smoothed in the full $SU(2)_L$ gauged
supergravity theory \cite{ChaVol1, ChaVol2}. 
We choose the ansatz
\begin{equation}
\label{ALrho}
A_L       = \frac{1}{g} \left( a(\rho) \,                d \theta \, \tau^1 + 
                               a(\rho) \, \sin \theta \, d \phi   \, \tau^2 + 
                                          \cos \theta \, d \phi   \, \tau^3 
                        \right), 
\end{equation}
where $\tau^l = \sigma^l/2$ are the $SU(2)$ generators. Obviously, $\delta_3$
is still a symmetry, and it can be verified that $\delta_2$ is still 
equivalent to the same singular gauge transformation, which is now manifestly
non--abelian,
\begin{equation}
\delta_2 A_L = \frac{1}{g} \, d \Lambda + i \, [\Lambda , A_L],
\end{equation} 
with
\begin{equation}
\Lambda = -\hepsilon \, \frac{\sin \phi}{\sin \theta} \, \tau^3.
\end{equation}

The appropriate supersymmetric solution is \cite{ChaVol1, ChaVol2}

\begin{eqnarray}
\label{e2hrho}
e^{2 h(\rho)} & = & \rho \coth (2 \rho) - \frac{\rho^2}{\sinh^2 (2 \rho)} 
                    - \frac{1}{4},     \\
\label{yrho}
y(\rho)       & = & \frac{8}{5} 
   \left( \varphi_0 + \half \log \left( 
         \frac{2 e^{h(\rho)}}{\sinh (2 \rho)} \right) \right), \\
\label{arho}
a(\rho)       & = & \frac{2 \rho}{\sinh (2 \rho)}.
\end{eqnarray}

The non--zero components of the gauge field strength are
\begin{eqnarray}
\label{FL1}
F_{L \rho \theta}^1 & = & -F_{L \theta \rho}^1 = \frac{1}{g} \, a'(\rho), \\
\label{FL2}
F_{L \rho \phi  }^2 & = & -F_{L \phi   \rho}^2 = 
                          \frac{1}{g} \sin \theta \, a'(\rho), \\
\label{FL3}
F_{L \theta \phi}^3 & = & -F_{L \phi \theta}^3 = 
                        -\frac{1}{g} \sin \theta \left(1 - a^2(\rho) \right).
\end{eqnarray}

This solution approaches the singular one for $\rho \rightarrow
\infty$.  In fact, if in the singular solution we substitute $\rho -
1/4$ instead of $\rho$, and change $\varphi_0$ appropriately, the
difference between the solutions involves only exponentially small
terms for large $\rho$.

Most importantly, this solution is regular at $\rho = 0$.
In fact, the geometry is becoming flat for $\rho \rightarrow 0$, 
as 
%(\ref{e2hrho},\ref{yrho})
\begin{eqnarray}
\label{e2hrhosmall}
e^{2 h(\rho)} & = & \rho^2 + O(\rho^4), \\
\label{yrhosmall}
y(\rho)       & = & \frac{8}{5} \varphi_0 + O(\rho^2).
\end{eqnarray}

\newsec{The Linearized Equation for the Scalars}

The equations of motion of the fields charged under $SU(2)_R$, that is,
the gauge vectors $A_R^r$ and the scalars $c_{lr}$ in the 
$\bf{9} = (\bf{3},{\bf 3})$
representation, derived from the Lagrangian (\ref{L7SO4}), are
\begin{eqnarray}
{\cal D}_\mu {\cal D}^\mu c_{lr} - 
\half e^{-y/2} (F_L^l F_R^r + F_L^l F_L^m c_{mr} + F_R^r F_R^s c_{ls})
    + O(c^2) & = & 0, \\
{\cal D}_\mu \left[ e^{-y/2} (F_R^{r \mu \nu} + 2 F_L^{l \mu \nu} c_{lr} + 
2 F_R^{s \mu \nu} c_{lr} c_{ls}) \right] + O(c^3) & = & 0,
\end{eqnarray}
where ${\cal D}_{\mu}$ is a metric covariant derivative as well as a
gauge covariant derivative.
We wish to solve those equations in perturbation theory around
the regular background described in the previous section.
We take the perturbation to be proportional to a small
parameter $\epsilon$, so that we can use the linearized form of the
equations,
\begin{eqnarray}
\label{ceom}
{\cal D}_\mu {\cal D}^\mu c_{lr} - 
\half e^{-y/2} (F_L^l F_R^r + F_L^l F_L^m c_{mr})
    + O(\epsilon^2) & = & 0, \\
\label{FReom}
{\cal D}_\mu \left[ e^{-y/2} (F_R^{r \mu \nu} + 2 F_L^{l \mu \nu} c_{lr}) 
                   \right] + O(\epsilon^2) & = & 0.
\end{eqnarray}
At first order in $\epsilon$, the fields $A_R^r$ can be taken as
three independent abelian gauge vectors, with the corresponding 
$F_R^r$ and covariant derivatives ${\cal D}_\mu$.   
Therefore, the equations above decouple into three sets of identical
equations for $r = 1,2,3$, and the general solution is a superposition of three
copies of the decoupled solutions. 
By a global $SU(2)_R$ transformation we may, if we wish, take $r = 3$
without loss of generality, and we will denote $A_R \equiv A_R^3$,
$F_R \equiv F_R^3$ and $c_l \equiv c_{l3}$, with
%We therefore simply denote by $A_R$
%and $F_R$ a $U(1)_R$ gauge vector and field strength, and also 
%define a field $c_l$ in
%the ${\bf 3}$ of $SU(2)_L$, with
\begin{equation}
{\cal D}_\mu c_l = \partial_\mu c_l + g \epsilon_{lmn} A_{L \mu}^m c_n.
\end{equation}

The simplest solutions will be those retaining the symmetries of the 
background. 
We begin by looking at the fields $c_l$.
Let us define $c = c_1 \tau^1 + c_2 \tau^2 + c_3 \tau^3$. 
Demanding the invariance of $c$ under the symmetry $\delta_3$ is obviously
equivalent to $\partial c / \partial \phi = 0$, that is, as a function 
of the angles, $c$ is dependent upon $\theta$ only:
\begin{equation}
\label{delta3ofc}
c(\theta , \phi) = f_3(\theta),  
\end{equation}
for an arbitrary function $f_3$.

Invariance of $c$ under the symmetry $\delta_2$ requires that the geometric 
transition
\begin{equation}
\delta_2 c = \frac{\partial c}{\partial \theta} \, \delta_2 \theta + 
             \frac{\partial c}{\partial \phi  } \, \delta_2 \phi 
\end{equation}
will be equivalent to the gauge transformation (\ref{Lambda})
\begin{equation}
\delta c = i \, [\Lambda, c] = \hepsilon \, \frac{\sin \phi}{\sin \theta} 
                               (c_1 \tau^2 - c_2 \tau^1),
\end{equation}
which gives
\begin{eqnarray}
\delta c_1 & = & - \hepsilon \, \frac{\sin \phi}{\sin \theta} c_2, \\
\delta c_2 & = & + \hepsilon \, \frac{\sin \phi}{\sin \theta} c_1, \\
\delta c_3 & = & 0.
\end{eqnarray}
This amounts to the partial differential equations
\begin{eqnarray}
D c_1 & = & - c_2, \\
D c_2 & = & + c_1, \\
D c_3 & = & 0,
\end{eqnarray}
with
\begin{equation}
D \equiv \sin \theta \cot \phi \, \frac{\partial}{\partial \theta} - 
         \cos \theta           \, \frac{\partial}{\partial \phi  }.
\end{equation}
Changing coordinates to
\begin{eqnarray}
u & = & \cos \theta \tan \phi, \\
v & = & \sin \theta \sin \phi,
\end{eqnarray}
we have
\begin{equation}
D = -(1 + u^2) \frac{\partial}{\partial u},
\end{equation}
and the solution can be seen to be  
\begin{eqnarray}
\label{delta2ofc1}
c_1 & = & \frac{1}{\sqrt{1 + u^{-2}}} f_2(v), \\
\label{delta2ofc2}
c_2 & = & \frac{1}{\sqrt{1 + u^{+2}}} f_2(v), \\
\label{delta2ofc3}
c_3 & = & \hat{f}_2(v), 
\end{eqnarray}
for arbitrary functions $f_2, \hat{f}_2$, or the exchanged solution 
$c_1 \rightarrow c_2, \,\, c_2 \rightarrow - c_1$.

\newsec{Solutions of the Linearized Equations} 

Demanding invariance under both $\delta_2$ 
(\ref{delta2ofc1},\ref{delta2ofc2},\ref{delta2ofc3}) 
and $\delta_3$ (\ref{delta3ofc})
is easily seen to 
lead to the vanishing of $c_1$ and $c_2$, and to $c_3$ being constant as 
a function of the angles, that is, being only a radial function. 
We will assume this in the rest of the paper.

In order for the field strength to respect this $SO(3)$ symmetry, it
can be seen that the gauge vector must be of the form 
%This assumption allows us to find a simple ansatz for the gauge vector,
\begin{eqnarray}
\label{ARsol}
A_R & = & \epsilon \cdot  \frac{\gamma}{g} \,           \cos \theta \,
            d \phi, \\
\label{FRsol}
F_R & = & \epsilon \cdot \left(-\frac{\gamma}{g}\right) \sin \theta \,
            d \theta \wedge d \phi, 
\end{eqnarray} 
where $\gamma$ depends only on the radial coordinate. 
This is nothing else than the form
(\ref{AL3sing}) of $A_L^3$ in the singular background.  
The equation of motion for $A_R$ (\ref{FReom}) can be seen to
be satisfied if and only if $\gamma$ is a constant.
Note that with only $c_3$ present, the only
non--vanishing components of 
$F_L^{l \mu \nu} c_l$ are for 
$\mu = \theta, \nu = \phi$ or vice versa. 
%Moreover, 
%$\sqrt{\det g_{\alpha \beta}} (F_R^{\mu \nu} + 2 F_L^{l \mu \nu} c_l)$ 
%is a radial function, and the angular covariant derivative vanishes.

However, this solution is singular at the origin $\rho = 0$. 
There, the contribution to the action from the sphere, parameterized by
$\theta$ and $\phi$, at a radial coordinate $\rho = \rho_0$, can be
seen from the Lagrangian (\ref{L7SO4}) and the metric approximation 
(\ref{e2hrhosmall},\ref{yrhosmall}) to behave like 
\begin{equation}
\int F_{R \mu \nu} F_R^{\mu \nu} \, \rho_0^2 \sin \theta \, d \theta \, d \phi
 \,\, \sim \,\, \rho_0^{-2},
\end{equation}
and the contribution from $\rho \ge \rho_0$ diverges as $\rho_0^{-1}$
when $\rho_0$ tends to zero. Therefore, this solution is 
physically unacceptable, and we will set $A_R=0$ from here on.

%Nevertheless, we will plug the solution (\ref{FRsol})
%into the equation of motion (\ref{ceom}). 
We remain
with the non--trivial equation of motion (\ref{ceom}) for $c_l$, 
which explicitly reads 
\begin{eqnarray}
c_3''(\rho) + \left(-\frac{5}{4} y'(\rho) + 2 h'(\rho)\right) c_3'(\rho) 
+ \mbox{} \hspace{95pt} & & \nonumber \\
\left(-2 e^{-2 h(\rho)} \, a^2(\rho) - 
      \frac{1}{2} \, e^{-4 h(\rho)} \, (1 - a^2(\rho))^2 \right) c_3(\rho) 
%+ \mbox{} \hspace{0pt} & & \nonumber 
%\\
%\left(-\frac{\epsilon \gamma}{2} \, e^{-4 h(\rho)} \, (1 - a^2(\rho)) \right) 
+ O(\epsilon^2) \hspace{0pt} & = & 0. 
\label{c3radialeq}
\end{eqnarray}
%this is an inhomogeneous equation, where the non--homogeneity results
%from the singular solution (\ref{ARsol},\ref{FRsol}). 
%
%For the time being we restrict ourselves to the homogeneous equation,
%i.e. taking $\gamma = 0$. 
Let us first of all study this equation in the limits 
$\rho \rightarrow 0$ and $\rho \rightarrow \infty$. In the former
limit, the space is flat (\ref{e2hrhosmall},\ref{yrhosmall})
and the gauge vector (\ref{ALrho}) tends to a constant,
\begin{equation}
\label{ALconst}
A_L(\rho) \approx \frac{1}{g} \left(                 d \theta \, \tau^1 + 
                                      \sin \theta \, d \phi   \, \tau^2 + 
                                      \cos \theta \, d \phi   \, \tau^3 
                              \right), 
\end{equation}
since from (\ref{arho}) we have
\begin{equation}
a(\rho) = 1 + O(\rho^2).
\end{equation}
Moreover, the field strength (\ref{FL1},\ref{FL2},\ref{FL3}) vanishes
in this limit,
\begin{equation}
\label{Fconst}
F_L \approx 0.
\end{equation}

Equation (\ref{c3radialeq}) turns in this limit into
\begin{equation}
c_3''(\rho) + \frac{2}{\rho} c_3'(\rho) - \frac{2}{\rho^2} c_3(\rho) 
            + O(\epsilon^2) = 0, 
\end{equation}
two of whose independent solutions are
\begin{eqnarray}
c_3(\rho) & = & \epsilon \rho^{-2} + O(\epsilon^2), \\
\label{c3rhoflat}
c_3(\rho) & = & \epsilon \rho      + O(\epsilon^2). 
\end{eqnarray}
A priori,
we can take a linear combination of those solutions with arbitrary
coefficients. 
However, we must demand that the coefficient of the first solution
vanishes, or else the solution diverges and is singular at the origin. 
At first glance, the second solution might also seem to be singular at
the origin, having a cusp like that of the absolute value function, and 
therefore being non--differentiable. However, gauge--invariant objects like
$c_3^2$ are smooth, and the gauge
covariant object appearing in the action
is the covariant derivative, having also a
contribution from the gauge vector $A_L$, which is singular at the
origin. Indeed, the constant gauge vector (\ref{ALconst}) is pure
gauge, in accordance with (\ref{Fconst}),
\begin{equation}
A_L = \frac{i}{g} \, G^{-1} \, d G,
\end{equation}
with 
\begin{equation}
G = e^{i \phi \tau^3} e^{i (\pi - \theta) \tau^1}.
\end{equation}
Therefore, we can move back to the gauge where $A_L = 0$, and then the
second solution (\ref{c3rhoflat}) looks as
\begin{equation}
c = G (\epsilon \rho \, \tau^3) G^{-1} = 
    \epsilon \left( y \tau^1 + x \tau^2 - z \tau^3 \right),
\end{equation}
which is perfectly regular, where we have changed from the polar
coordinates $\rho, \theta, \phi$ to the Cartesian ones $x, y, z$.

We now wish to look at the limit $\rho \rightarrow \infty$. There, we
can use the singular solution (\ref{e2hrhosing},\ref{yrhosing},\ref{AL3sing})
having $a(\rho) = 0$. Equation (\ref{c3radialeq}) becomes 
\begin{equation}
c_3''(\rho) + \left( 2 + \half \rho^{-1} \right) c_3'(\rho) 
            - \half \rho^{-2} c_3(\rho) + O(\epsilon^2) = 0,
\end{equation}
two of whose independent solutions are
\begin{eqnarray}
\label{c3rhoinftynorm}
c_3(\rho) & = & \epsilon \left( e^{-2 \rho} \rho^{-1/2} \right) 
                + O(\epsilon^2), \\
\label{c3rhoinftynonnorm}
c_3(\rho) & = & \epsilon \left( 1 - \frac{\sqrt{2 \pi}}{4} e^{-2 \rho}
                         \rho^{-1/2} \erfi \left(\sqrt{2 \rho}\right) \right) 
                + O(\epsilon^2), 
\end{eqnarray}
where $\erfi$ is an imaginary version of the error function, 
$\erfi(z) = \erf(i z) / i$, obeying
$\erfi'(z) = (2/\sqrt{\pi}) \exp(+ z^2)$.
The first solution is exponentially small at $\rho \rightarrow \infty$, 
and is therefore normalizable, while the second behaves as 
$ 1 - (1/4) \rho^{-1} + O(\rho^{-2})$ and is non--normalizable. 
The second solution corresponds to a perturbation of the Lagrangian by
the appropriate operator $\epsilon {\cal O}$, 
while the first corresponds to a vacuum
expectation value (VEV) of the operator. 
However, we do not expect to have a physically acceptable solution 
corresponding to a VEV, as the dual field theory does not have a moduli space.
Indeed, we will see that the full solution corresponding to having only 
a VEV is singular at the origin and therefore unacceptable. 
%Note that one can always add the normalizable solution to the 
%non--normalizable one, and the correct
%coefficient needed to be added of the former can't be
%determined from the behaviour at the boundary $\rho = \infty$. 
%Rather, it is determined from 
Global regularity of the full solution, or in
other words the boundary condition we have described at $\rho = 0$,
will determine a particular linear combination of (\ref{c3rhoinftynorm})
and (\ref{c3rhoinftynonnorm}).

Equation (\ref{c3radialeq})
%,still looking at the homogeneous level, 
can be solved analytically in the
whole region $0 \leq \rho < \infty$. Defining
\begin{equation}
q(\rho) = \sqrt{4 \rho \sinh(4 \rho) - \cosh(4 \rho) - 8 \rho^2 + 1},
\end{equation}
we have as two independent solutions
\begin{eqnarray}
\label{c3rhobad}
c_3(\rho) & = & \epsilon \left( \frac{1}{q(\rho)} \right) 
                + O(\epsilon^2), \\
\label{c3rhogood}
c_3(\rho) & = & \epsilon \left( \frac{\int_{0}^{\rho} q(s) \, ds}{q(\rho)} \right) 
                + O(\epsilon^2). 
\end{eqnarray}

For small values of $\rho$ we have 
$q(\rho) = 4 \sqrt{2} \, \rho^2 + O(\rho^4)$.
Therefore, the first solution behaves at $\rho \rightarrow 0$ as
$\rho^{-2}$ and is unacceptable, while the second behaves as 
$(1/3) \rho + O(\rho^3)$ so it is the one we're looking for. 

\begin{figure}[h!]
\begin{center}
\resizebox{0.6\textwidth}{!}{\includegraphics{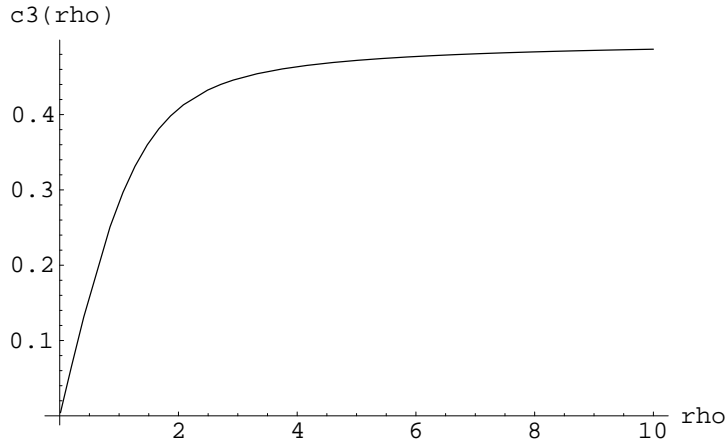}}
\end{center}
\caption{The graph of $c_3(\rho)/\epsilon$ as given in (\ref{c3rhogood}).}
\label{fig:c3rho}
\end{figure}

As $\rho \rightarrow \infty$, we have
\begin{equation}
q(\rho) = \sqrt{2 \rho - \half} \cdot e^{2 \rho} \cdot 
          \left( 1 + O(\rho^2 e^{-4 \rho}) \right).
\end{equation}
If we substitute $q(\rho) \approx \sqrt{2 \rho} \, e^{2 \rho}$ into
(\ref{c3rhogood}), we get exactly half of (\ref{c3rhoinftynonnorm}), while
clearly the same substitution in (\ref{c3rhobad}) gives $1/\sqrt{2}$
of (\ref{c3rhoinftynorm}). So, we see again that the solution including
the full smoothed background
behaves at infinity like the solution including the singular background, 
with a ``lagging'' of
$\rho$ by $1/4$, up to exponentially small corrections.

%We now write also particular solutions $\tilde{c}_3$ of the full 
%inhomogeneous equation (\ref{c3radialeq}).
%For the singular background, this solution can be taken simply as a
%constant,
%\begin{equation}
%\label{c3rhoinhomsing}
%\tilde{c}_3(\rho) = - \epsilon \, \gamma 
%\end{equation}
%For the smooth background, we may take  
%\begin{equation}
%\label{c3rhoinhom}
%\tilde{c}_3(\rho) = \epsilon \, \gamma 
%\left( \frac{\int_0^r (8 s - 2 \sinh(4 s)) / q(s) \, ds}{q(\rho)} \right)
%\end{equation} 
%This solution is even in $\rho$ and is regular at the origin: 
%$\tilde{c}_3(\rho) \approx \epsilon \, \gamma 
%\left( -1/3 + (4/45) \rho^2 \right)$. 
%At infinity, it behaves as 
%$\tilde{c}_3(\rho) \approx \epsilon \, \gamma 
%\left( -(1/4) \rho^{-1} \right)$.

To recapitulate, we found three modes of small perturbations around
the Maldacena--Nu\~nez solution: only one (\ref{c3rhogood}) is a
smooth, physically acceptable radial scalar mode $c_3(\rho)$. 
The second mode (\ref{c3rhobad}) is singular in the IR, and the third,
described by
(\ref{ARsol}) with an appropriate solution for $c_3(\rho)$,
involves a singular gauge vector $A_R$.
% accompanied by $\tilde{c}_3(\rho)$. 

%\begin{figure}[h!]
%\begin{center}
%\resizebox{0.6\textwidth}{!}{\includegraphics{c3rhoinhom.eps}}
%\end{center}
%\caption{The graph of $\tilde{c}_3(\rho)$ as given in (\ref{c3rhoinhom})}
%\label{fig:c3rhoinhom}
%\end{figure}

\newsec{The Ten--Dimensional Solution}
The seven--dimensional solution can be lifted up to ten dimensions
using the expressions in \cite{CvLuPo}. In that paper, the three
dimensional sphere $S^3$ used for the lifting is parameterized by $\mu^a$,
where $a = 0,1,2,3$ is an $SO(4)$ vector index, and
\begin{equation}
\mu^a \mu^a = 1.
\end{equation}
The covariant derivative of $\mu^a$ is defined as 
\begin{equation}
{\cal D} \mu^a = d \mu^a + {\hat g} A_{(1)}^{ab} \mu^b.
\end{equation}
Maldacena and Nu\~{n}ez \cite{MalNun} employ the identification of
$S^3$ with the group manifold of $SU(2)$, and use instead three Euler angles, 
which we will denote by $\tpsi, \ttheta, \tphi$. 
Their solution can be written in terms of the three one--forms $w^l$ defined
by
\begin{eqnarray}
w^1 + i w^2 & = & e^{-i \tpsi} (d \ttheta + i \sin \ttheta d \tphi), \\
w^3         & = &               d \tpsi   +   \cos \ttheta d \tphi.
\end{eqnarray}
The $w^l$ have, in our conventions, an $SU(2)_L$ index $l=1,2,3$, and are
invariant under $SU(2)_R$. 

The translation between these two parameterizations is given by 
\begin{eqnarray}
\label{mu0}
\mu^0 & = & +\cos (\frac{\tpsi + \tphi}{2}) \, \cos (\frac{\ttheta}{2}), \\
\label{mu1}
\mu^1 & = & -\cos (\frac{\tpsi - \tphi}{2}) \, \sin (\frac{\ttheta}{2}), \\
\label{mu2}
\mu^2 & = & +\sin (\frac{\tpsi - \tphi}{2}) \, \sin (\frac{\ttheta}{2}), \\
\label{mu3}
\mu^3 & = & -\sin (\frac{\tpsi + \tphi}{2}) \, \cos (\frac{\ttheta}{2}), 
\end{eqnarray}
and we have
\begin{equation}
w^l = \frac{4}{i} \, (\alpha_L^l)_{ab} \, \mu^a \, d \mu^b.
\end{equation} 
Those one--forms will appear in the formulae through the combinations
\begin{equation}
\label{hw}
\hw^l \equiv w^l - g A_L^l. 
\end{equation} 

We also have to remember the relation between $g$ and $\hat g$
explained in section
\ref{sec:lagrangian}.
%the $g$ of \cite{CvLuPo} by $-(1/\sqrt{2}) g$. 
Because in our solution the $SU(2)_R$ gauge fields vanish, we find the
simple relation
\begin{equation}
{\cal D} \mu^a = i \, (\alpha_L^l)_{ab} \, \mu^b \hw^l.
\end{equation}
Using the anti--commutators (\ref{ALanticom}) from appendix A, this
gives, in particular,
\begin{equation}
{\cal D} \mu^a {\cal D} \mu^a = \frac{1}{4} \hw^l \hw^l.  
\end{equation}

The radial solution of the scalar equation which we found in the last section
is $c_3(\rho)$ which, without loss of generality, can be taken to be
the field $c_{33}(\rho)$.
The first order expression in $\epsilon$ for the scalars we need to
maintain is then, by
(\ref{tildeT},\ref{trQ}), 
\begin{equation}
\tilde{T}_{ab} = \delta_{ab} + c_3 (X_{33})_{ab} +O(\epsilon^2).
\end{equation}
The lifting up expressions of \cite{CvLuPo} involve, among other
things, the combination $\tilde{T}_{ab} \mu^a \mu^b$. Using the
explicit form of $X_{33}$ from appendix A, and
(\ref{mu0}--\ref{mu3}), we get 
\begin{eqnarray}
\tilde{T}_{ab} \mu^a \mu^b & = & 
1 + c_3 \left( (\mu^0)^2 - (\mu^1)^2 - (\mu^2)^2 + (\mu^3)^2 \right)
+ O(\epsilon^2) \nonumber \\
& = & 1 + c_3 \cos \ttheta + O(\epsilon^2).
\end{eqnarray} 
Another expression which appears in the formulae is
\begin{eqnarray}
Z & \equiv & (X_{33})_{ab} {\cal D} \mu^a {\cal D} \mu^b = 
 ({\cal D} \mu^0)^2 - ({\cal D} \mu^1)^2 - 
 ({\cal D} \mu^2)^2 + ({\cal D} \mu^3)^2   \nonumber \\     
  & = & \frac{1}{4} 
\left( 
2 \sin \ttheta \, (\sin \tpsi \, \hw^1 + \cos \tpsi \, \hw^2)
                  \, \hw^3 -
  \cos \ttheta \, ((\hw^1)^2 + (\hw^2)^2 - (\hw^3)^2)           
\right).
\end{eqnarray}
There is an important point to notice here. In the asymptotic region 
$\rho \rightarrow \infty$, corresponding to the UV, the 
seven--dimensional solution resembles the singular one, having only an
$A_L^3$ gauge field.
Therefore we may take 
$\hw^l = w^l$ for $l=1,2$ in that region, and it is easily seen that
$Z$ involves the coordinate $\tpsi$ only through $d \tpsi$. 
For example, 
\begin{equation}
\sin \tpsi \, w^1 + \cos \tpsi \, w^2 = \sin \ttheta \, d \tphi.
\end{equation}
In other
words, $Z$ is asymptotically invariant under a constant shift in $\tpsi$.

Other expressions for various differential forms 
which we will find useful are 
\begin{eqnarray}
\Xi_2 & \equiv &
\sin \ttheta \, (\sin \tpsi \, \hw^1 + \cos \tpsi \, \hw^2) \wedge \hw^3, \\
\Psi_3 & \equiv & 
 -(\sin \tpsi \, g F_L^1 + \cos \tpsi \, g F_L^2) \wedge \hw^3 +
  g F_L^3 \wedge (\sin \tpsi \, \hw^1 + \cos \tpsi \, \hw^2), \\
\Pi_3 & \equiv & 
    \sin \ttheta \, (\sin \tpsi \, g A_L^1 + \cos \tpsi \, g A_L^2) 
      \wedge \hw^1 \wedge \hw^2 + \mbox{} \nonumber \\
& & \cos \ttheta \, (g A_L^2 \wedge \hw^1 - g A_L^1 \wedge \hw^2) \wedge \hw^3. 
\end{eqnarray}
Those expressions are also asymptotically invariant under a constant shift in
$\tpsi$ as, by (\ref{ALrho},\ref{arho},\ref{FL1},\ref{FL2}), 
$A_L^1 , A_L^2$ and also $F_L^1 , F_L^2$ are asymptotically zero.

We are now in the position to write explicitly the ten--dimensional
Einstein frame metric, the dilaton, and the Neveu--Schwarz
three--form. 
We find respectively 
\begin{eqnarray}
d s_{10}^2 & = & e^{-5 y / 16} \left[ 
  (1 + \frac{1}{4} c_3 \cos \ttheta) d s_{7,\mbox{st}}^2 +
\frac{2}{g^2} (1 - \frac{3}{4} c_3 \cos \ttheta) 
              (\frac{1}{4} \hw^l \hw^l - c_3 Z) \right] + \mbox{} 
\nonumber \\ \label{metric10d}
& & O(\epsilon^2), \\
\varphi & = & 
  \frac{5}{8} y - \frac{1}{2} c_3 \cos \ttheta + O(\epsilon^2) 
\nonumber \\ \label{dilaton10d} 
        & = &  
  \varphi_0 +  \half \log \left(\frac{2 e^{h(\rho)}}{\sinh (2 \rho)} \right) 
  - \frac{1}{2} c_3 \cos \ttheta + O(\epsilon^2), \\
\HNS & = & \frac{2}{g^2} 
\left[-\frac{1}{4} (1 - 2 c_3 \cos \ttheta) \hw^1 \wedge \hw^2 \wedge \hw^3 +
       \frac{1}{4} g F_L^l \wedge \hw^l +
       \frac{1}{4} c_3 \sin \ttheta \, \Psi_3 - \mbox{} \right. 
\nonumber \\ \label{flux10d} 
 & & \hspace{24pt} \left.   
       \frac{1}{4} \, c_3 \, \Pi_3 - 
       \frac{1}{4 \sqrt{2}} \, \Xi_2 \wedge d c_3 
\right] + O(\epsilon^2). 
\end{eqnarray}
In those formulae we have used the Einstein summation convention, 
and in our radial case we have $d c_3 = c_3'(\rho) \, d \rho$.
Note that the combination $c_3 \cos \ttheta$, which appears above in
the expressions for fields which are $SU(2)_R$--invariant, is indeed
covariant under $SU(2)_R$.

Those formulae reduce to the solution of \cite{MalNun} when the
scalar $c_3(\rho)$ is turned off. In our formulation, however, an
explicit factor of $g$ accompanies each $A_L^l$ or $F_L^l$, as in
(\ref{hw}), because of the prefactor $1/g$ in the ansatz
(\ref{ALrho}). 
The additional terms in our solution are
proportional to the scalar $c_3(\rho)$, given by (\ref{c3rhogood}),
and thus to the small 
parameter $\epsilon$ in which we expand.

It is interesting to note that in the string frame, the 
seven--dimensional part of the ten--dimensional metric does not contain a
contribution linear in $\epsilon$, 
\begin{equation}
\label{smetric10d}
d s_{10,\mbox{st}}^2  =  
d s_{7, \mbox{st}}^2 +
\frac{2}{g^2} (1 - c_3 \cos \ttheta) (\frac{1}{4} \hw^l \hw^l - c_3 Z) + 
O(\epsilon^2).
\end{equation}

\newsec{The S--dual Background and the String Tension}

%In order to describe the dynamics of the SYM field theory in terms of
%the fundamental string, we have
In order to find the tension of the Wilson line in the field theory it
will be convenient
to S--dualize the background, or to wrap D5 branes on the two--sphere
instead of NS5 branes. The string frame metric S--dual to 
(\ref{smetric10d})
%, in terms of the original metric, 
is  
\begin{equation}
d s_{10,\mbox{st} , D}^2  = e^{\varphi_D} d s_{10,\mbox{st}}^2,
\end{equation}
where the S--dual dilaton is given by
%to (\ref{dilaton10d}), is
\begin{equation} 
\varphi_D = 
  \varphi_{D , 0} -  
 \half \log \left(\frac{2 e^{h(\rho)}}{\sinh (2 \rho)} \right) 
  + \frac{1}{2} c_3 \cos \ttheta + O(\epsilon^2), 
\end{equation}
and the Neveu--Schwarz field $\HNS$ turns into the S--dual
Ramond--Ramond field $\HRR$. 

We may probe the YM theory by heavy external quarks.
The fluxtube between a quark anti--quark pair is described in the S-dual
gravitational dual by a
fundamental string, lying on a geodesic of the dual supergravity
background, and asymptoting to $\rho = \infty$ at both ends \cite{Mal2,ReYe}. 
The confining potential is given by the renormalized mass of such a string.
For large separations, the string will tend to be stretched where the
four--dimensional metric is minimal, the string tension will be
given in terms of the metric at 
that minimum, and the corrections to the linear potential are
very small \cite{KSS2}. In our case, the minimum is obtained where
$\varphi_D$ attains the minimal value. Before the perturbation,
$\varphi_D$ had a quadratic minimum at the origin $\rho = 0$. 
The perturbation by $c_3$ adds a function (\ref{c3rhogood}) which is 
linear in $\rho$ near the origin and is proportional to the small
parameter $\epsilon$ times $\cos \ttheta$. Taking $\epsilon > 0$ the
minimum will be attained for $\cos \ttheta = -1$, that is at the south
pole $\ttheta = \pi$ of the three--sphere, and
for $\rho_\tmin = \frac{3}{16} \epsilon$. However, the value of the
metric at the minimum arising from this shift is only corrected at
quadratic order in $\epsilon$, and it
%its value, however, 
might be affected also by terms in $\varphi_D$
which are quadratic in $\epsilon$. All we can say is, therefore, that 
$\varphi_{D , \tmin} \equiv \varphi_D(\rho_\tmin) = 
\varphi_{D , 0} + O(\epsilon^2)$.
The string tension 
$T_{\mbox{st}} = e^{\varphi_{D , \tmin}} / 2 \pi \alpha'$
may then also be corrected relative to the unperturbed
value, with a correction which is quadratic in the strength of the
(small) perturbation.

Other properties of the dual gauge dynamics like  the L\"uscher term, the
broadening of the flux tube, stringy
corrections to the intercept, 't Hooft loops, baryonic configurations,
the gaugino condensate and the domain wall tensions will behave as in the
Maldacena--Nu\~nez case \cite{MalNun,Loewy:2001pq,Apreda:2001qb}. 
The $U(1)$ R--symmetry corresponds in the ten--dimensional picture to
a constant shift of the $\tpsi$ coordinate \cite{MalNun}. 
As shown in the previous section, this is a symmetry of the asymptotic
form of our solution for large $\rho$. 
The breaking of this symmetry by worldsheet instantons will also
remain as in \cite{MalNun}, except for a small change in the
shape of the dominant worldsheet configurations.

\newsec{Supersymmetry and its Breaking}

In this section we exhibit explicitly the supersymmetry of the 
Maldacena--Nu\~nez solution, and we show that it is broken by the
deformation. It is sufficient for us to work in the
singular solution, which is simpler, because we break supersymmetry in
the UV, where the two solutions are similar.
Supersymmetry acts locally, so
the UV behaviour should not be affected by the smoothing 
in the IR region.
We work in ten dimensions, 
in the context of type IIB supergravity, although the solution
is contained in the sector of type I fields. The four--dimensional
Minkowski space $x^\mu , \mu = 0, 1, 2, 3$ plays no role in the 
considerations, and therefore we may work solely with the six coordinates 
$\rho, \theta, \phi, \ttheta, \tphi, \tpsi$.
We choose an eight--dimensional Majorana representation for the 
flat Euclidean $SO(6)$ Clifford algebra, that is, purely imaginary 
gamma matrices 
$\gamma^i , i = 1 , \ldots , 6$ satisfying 
$\{\gamma^i , \gamma^j\} = 2 \delta^{i j}$. 
The gamma matrices for our 
six--dimensional curved space will be denoted by 
$\Gamma^M , M = 1, \ldots , 6$,
where $\Gamma^M = e^{i M} \gamma^i$, with $e^{i M}$ being the
vielbein, $e^i_M e^i_N = g_{M N}$.

The background is bosonic, and therefore the supersymmetry
transformations of the bosons vanish, and those of the fermions
involve only the bosonic fields. Remember also that our background
contains no five--form field. In our conventions, the supersymmetry
transformations with a spinor parameter $\eta$ for the dilatino and gravitino
read, respectively,
\begin{eqnarray}
\label{deltalambda}
\delta \lambda & = & 
  \frac{i}{ 2} \Gamma^M \partial_M \varphi \, \eta^* -
  \frac{i}{24} e^{-\varphi/2} \Gamma^{M N P} H_{M N P} \, \eta, \\
\label{deltapsiM}
\delta \psi_M  & = & 
  {\cal D}_M \eta + 
  \frac{1}{96} e^{-\varphi/2} (  \Gamma_M^{\hspace{8pt} N P Q} H_{N P Q} - 
                               9 \Gamma^{N P}     H_{M N P}   ) \eta^*.
\end{eqnarray}
The covariant derivative of the spinor is 
${\cal D}_M = \partial_M + \frac{i}{2} \omega_M^{i j} \Sigma^{i j}$, where 
$\omega_M^{i j}$ is the spin connection, and 
$\Sigma^{i j} = -\frac{i}{4} [\gamma^i , \gamma^j]$ are the generators
of rotations in the spinor representation. The multiple index gamma
matrices are defined as antisymmetrized products of unit weight, 
that is, for example, 
$\Gamma^{M N} = \frac{1}{2!} (\Gamma^M \Gamma^N - \Gamma^N \Gamma^M)$.
%We work in a Majorana representation, where the 
Since our
gamma matrices are imaginary, 
%and therefore 
the supersymmetry
transformations (\ref{deltalambda},\ref{deltapsiM}) are real
operators, acting both on real and on imaginary spinors. 
Hence the equations for a complex spinor $\eta$ decompose into
equations for its real part and for its imaginary part, and we can
work separately with the two cases.  

The unperturbed singular solution is rather inelaborate. 
The dilaton and Neveu--Schwarz three--form are
\begin{eqnarray}
\label{singphi}
\varphi & = & \varphi_0 - \rho + \frac{1}{4} \log \rho, \\
\HNS    & = & \frac{2}{g^2} \left(-\frac{1}{4}\right)
  \left[\sin \ttheta 
        (                d \ttheta \wedge d \tphi   \wedge d \tpsi -
         \cos  \theta \, d  \phi   \wedge d \ttheta \wedge d \tphi   )
         + \mbox{} \right. \nonumber \\ & & \hspace{57pt} \left.       
         \sin  \theta        
        (                d  \theta \wedge d  \phi   \wedge d \tpsi +
         \cos \ttheta \, d  \theta \wedge d  \phi   \wedge d \tphi   )
  \hspace{16pt} \right].  
\end{eqnarray}  
This considerably simplifies the supersymmetry equations. 
Choosing $\eta$ real, $\eta^* = \eta$, 
and writing $\delta \lambda = M^{(0)} \eta$, 
we find that the matrix $M^{(0)}$ has a two--dimensional kernel,
$\Ker M^{(0)} = \Span \{\zeta_1 , \zeta_2\}$ 
(see appendix C for some explicit details). 
Specializing to this subspace, i.e.\ writing
\begin{equation}
\eta = B_1 \zeta_1 + B_2 \zeta_2,
\end{equation} 
the equations for the gravitino, $\delta \psi_M = 0$, are
satisfied for $M = \theta, \phi, \ttheta, \tphi$ with vanishing
ordinary derivative, so the solution is consistent with no dependence
of $\eta$ on those coordinates. 
From the equation $\delta \psi_\rho = 0$ we get that
\begin{eqnarray}
\partial_\rho B_1 - 
\frac{1}{8} \left(1 - \frac{1}{4 \rho}\right) B_1 & = & 0, \\ 
\partial_\rho B_2 - 
\frac{1}{8} \left(1 - \frac{1}{4 \rho}\right) B_2 & = & 0,
\end{eqnarray}
from which we extract the radial dependence of the spinor: 
both $B_1$ and $B_2$ are proportional to $\rho^{-1/32} e^{\rho/8}$.

Finally, from the equation $\delta \psi_{\tilde{\psi}} = 0$ we get 
\begin{eqnarray}
\partial_{\tilde{\psi}} B_1 + \half B_2 & = & 0, \\ 
\partial_{\tilde{\psi}} B_2 - \half B_1 & = & 0. 
\end{eqnarray}
The solution looks like 
$B_1 \sim \cos (\tpsi/2) \,,\, B_2 \sim \sin (\tpsi/2)$
up to an overall factor and a constant shift in $\tpsi$. We find
that the spinors have the appropriate charge under the 
$U(1)$ R--symmetry taking
$\tpsi \rightarrow \tpsi + \delta \tpsi$, and that they acquire a phase of
$-1$ when $\tpsi \rightarrow \tpsi + 2 \pi$. 
In particular \cite{MalNun}, the periodicity of $\tpsi$ is
$\tpsi \equiv \tpsi + 4 \pi$. 

All in all, we get that the remaining supersymmetry is generated by
the spinors 
\begin{eqnarray}
\label{eta}
\eta & = & E \rho^{-1/32} e^{\rho/8} 
  \left(\cos \frac{\tpsi-\tpsi_0}{2} \, \zeta_1 + 
        \sin \frac{\tpsi-\tpsi_0}{2} \, \zeta_2   \right),
\end{eqnarray}
where $E$ and $\tpsi_0$ are arbitrary constants.
Choosing $\eta$ to be imaginary, $\eta^* = - \eta$, we still get a
two--dimensional solution for the dilatino equation, but this solution
is not consistent with the gravitino equations. Therefore we are left
with one eighth of type IIB supersymmetry, or one fourth of type I
in ten dimensions, which corresponds to ${\cal N} = 1$ in four dimensions.

Next,
we wish to show that the perturbation by the radial scalar (in the
seven--dimensional language) indeed totally breaks supersymmetry. To this
end it is enough to show that the dilatino equation cannot continue to
be satisfied in the presence of the perturbing field $c_3$. The supersymmetry
transformations of the dilatino and gravitino are modified, and the spinor
$\eta$ also might change, 
$\eta = \eta^{(0)} + \epsilon \eta^{(1)} + O(\epsilon^2)$, where
$\eta^{(0)}$ is the real solution (\ref{eta}) to the unperturbed equation,
$M^{(0)} \eta^{(0)} = 0$. 
If $\eta^{(1)}$ has an imaginary part, then the imaginary parts of the
$O(\epsilon)$ supersymmetry equations are identical to the original
equations on imaginary spinors, which we found to have no
solutions. Therefore we can assume that $\eta^{(1)}$ is also real.

The supersymmetry
transformation of the dilatino is changed to $\delta \lambda = M \eta$
where $M = M^{(0)} + \epsilon M^{(1)} + O(\epsilon^2)$. 
Positing that supersymmetry might be
conserved, we must have 
\begin{eqnarray}
0 & = & \delta \lambda = M \eta = 
\left(M^{(0)} + \epsilon M^{(1)}\right) (\eta^{(0)} + \epsilon \eta^{(1)})   
              + O(\epsilon^2) \nonumber \\
  & = & \epsilon \left( M^{(1)} \eta^{(0)} + M^{(0)} \eta^{(1)}\right)
              + O(\epsilon^2),
\end{eqnarray}   
or
\begin{equation}
M^{(0)} \eta^{(1)} = - M^{(1)} \eta^{(0)},
\end{equation}
which can be stated as $M^{(1)} \eta^{(0)} \in \Image M^{(0)}$,
or finally as $M^{(1)} \eta^{(0)} \bot \left(\Image M^{(0)}\right)^\bot$.
Checking this explicitly (see appendix C) yields the condition
\begin{equation}
\label{susycondition}
c_3'(\rho) + \frac{1 + 4 \rho}{2 \rho} c_3(\rho) = 0.
\end{equation}
Remember that this is a necessary condition for having some remaining
supersymmetry, coming only from the dilatino equation.

Taking the non--normalizable radial scalar mode 
(\ref{c3rhoinftynonnorm}) as the perturbation of the singular background, 
we see that (\ref{susycondition}) is not satisfied. Therefore, this
perturbation, which is dual to perturbing the Lagrangian in the field
theory, breaks supersymmetry.

It is interesting to note that 
(\ref{susycondition}) is satisfied for the normalizable 
radial scalar mode (\ref{c3rhoinftynorm}) of the same background.
This might suggest that supersymmetry is conserved for such a
deformation of the solution 
(although the gravitino equations should also be checked
in order to verify this). This is natural from the dual viewpoint of
the field theory, where such a mode corresponds to having a VEV of 
the operator ${\cal O}$, which does not break
supersymmetry. However, the physical relevance of such a mode is
not clear since the true, smooth, solution has no normalizable mode, and
the field theory has no moduli space.

\newsec{Summary and Discussion}

We began by expanding the seven--dimensional 
$SO(4) \equiv SU(2)_L \times SU(2)_R$ gauged 
supergravity Lagrangian (\ref{L7SO4}) up to second order in the
$\bf{9} = (\bf{3},{\bf 3})$ scalar fields $c_{lr}$. 
We reviewed the singular solution 
(\ref{e2hrhosing},\ref{yrhosing},\ref{AL3sing}) and the smooth 
Chamseddine--Volkov solution (\ref{e2hrho},\ref{yrho},\ref{arho}) of
that Lagrangian 
%consistently reduced to the gauged $SU(2)_L$ case without
%the scalars, 
and studied their symmetries. We then took the
scalars to be proportional to the small parameter $\epsilon$, and
wrote their linearized (that is, $O(\epsilon)$) equations of motion.
Those equations split into three identical copies, indexed by $r=1,2,3$,
of decoupled equations, 
and any superposition of such solutions $c_{lr}$ can be brought 
to the form $c_{l3}$ by a suitable $SU(2)_R$ rotation.
We therefore wrote those equations (\ref{ceom},\ref{FReom}) in terms of a
single
$SU(2)_L$ triplet of scalars $c_l$. Working around the aforementioned
singular and smooth solutions, and keeping their symmetries, we found that
the perturbation
must be a radial field $c_3(\rho)$ obeying the equation of motion
(\ref{c3radialeq}). 

Around the singular solution we found three independent
modes. The first (\ref{c3rhoinftynorm}) is normalizable and
corresponds to a VEV of the operator 
${\cal O} = 
\tr \left( \Phi_1^2 - \Phi_2^2 - \Phi_3^2 + \Phi_4^2 \right)$
 in the dual six--dimensional field theory. 
The second (\ref{c3rhoinftynonnorm}) is non--normalizable and
corresponds to a perturbation by that operator, and the third mode 
%(\ref{c3rhoinhomsing}) 
includes a singular gauge field $A_R$
given by (\ref{ARsol}). Around the physical, smooth solution, 
%we have
%also identified such three modes, which are, respectively, 
%(\ref{c3rhobad}), (\ref{c3rhogood}) and (\ref{c3rhoinhom}). Only the
we found that only the
second of these modes (\ref{c3rhogood}),
corresponding to the perturbation by $\epsilon {\cal O}$,
is smooth and physically acceptable, 
the other two being singular in the IR. 
Indeed, the dual field theory has
no moduli space and therefore no VEV is possible. 
Then, we raised the seven--dimensional solution to a ten--dimensional one 
(\ref{metric10d},\ref{dilaton10d},\ref{flux10d}). This is a stable
solution as it is dual to a field theory exhibiting a mass gap.  

The most important remaining issue is to find the solution to order
$\epsilon^2$. This involves finding the back--reaction of the perturbation
$c_3(\rho)$ on the fields of the Chamseddine--Volkov
background in seven dimensions, and raising it again to ten dimensions.
In order to find the back--reaction we should retain in the Lagrangian 
(\ref{L7SO4}) all the $O(\epsilon^2)$ terms. 
Working with a solution and a perturbation which both do not involve 
the $SU(2)_R$ gauge vectors $A_R^r$ to first order in $\epsilon$,
%(that is, taking $\gamma = 0$ 
%in (\ref{ARsol})), 
as we must for the physical perturbation, 
those gauge vectors can be neglected also in the back--reaction, 
because they do not contribute to the Lagrangian to order $\epsilon^2$. 
In particular
\begin{equation}
{\cal D}_\mu c_l = \partial_\mu c_l + g \epsilon_{lmn} A_{L \mu}^m c_n, 
\end{equation}
and the relevant Lagrangian takes the form
\begin{eqnarray}
\label{L7backreaction}
L_7 & = & R - \frac{5}{16} \partial_\mu y \partial^\mu y 
            - {\cal D}_\mu c_l {\cal D}^\mu c_l  \nonumber \\
    &   &   -\frac{1}{4} e^{-y/2} (F_L^l F_L^l + 2 F_L^l F_L^m c_l c_m)
            + 2 g^2 e^{y/2} + O(\epsilon^3).
\end{eqnarray}
The equations of motions arising from this Lagrangian at second order
in $\epsilon$ are
cumbersome coupled equations, but can be solved in principle. 

The back--reaction computation is necessary for computing the energy of the
vacuum (such a computation was carried out in a different context in
\cite{GuTsVo}). Since supersymmetry is broken at order $\epsilon$, we
expect to have a non--zero vacuum energy of order $\epsilon^2$; note that
since the deformation is in the $\bf 3$ representation of $SU(2)_R$, all
singlets of $SU(2)_R$ (such as the vacuum energy) must depend at least
quadratically on $\epsilon$.
%, because the
%$SU(2)_R$ charged fields appear at least quadratically in
%(\ref{L7SO4}), as the Lagrangian can be consistently truncated not to
%contain them. We expect the energy to be non--zero, as supersymmetry is broken
%to first order in $\epsilon$. In fact, all the physical effects of the
%perturbation are expected to behave at least quadratically in
%$\epsilon$, as they do not depend on the sign of $\epsilon$. 
In particular, we saw this in the computation of the string tension.
% in terms 
%of the dilaton $\varphi_D$ of the S--dual Maldacena--Nu\~nez background. 
It would be interesting to complete this calculation, as well as to find 
the corrections to the glueball masses and to the domain wall tensions (the
domain walls are no longer BPS in the deformed solution).

In section 8, we explicitly demonstrated that supersymmetry is broken in ten
dimensions by the small perturbation of the background. 
It would be nice to be able to check this directly in seven
dimensions. Such a computation would be somewhat more direct and simple, and
would presumably allow to deal in a feasible manner even with the 
smooth background. 
The supersymmetry transformations for the seven--dimensional $SO(4)$ 
gauged supergravity including the nonet of scalars can be extracted
from \cite{Salam:1983fa}.

%Regrettably, the supersymmetry breaking we have found is
%rather dull, as the classical $U(1)$ R--symmetry is maintained. As we
Even though the perturbation we described breaks supersymmetry, it does
not break the $U(1)$ R--symmetry, so many of the qualitative features
of the solution of \cite{MalNun} persist. In particular,
through the anomalous and spontaneous breaking of this
symmetry we are still left with $N$ equivalent vacua, which are
permuted by the action of $\IZ_N$. These vacua, in the language of
the YM theory, differ only in the phase of the gaugino bilinear
condensate. The linearized breaking of supersymmetry involves only
the scalars $\Phi_a$,
%The breaking of supersymmetry is achieved through the
%sector of the Kaluza--Klein modes of the squashed three--sphere, 
and it does not couple directly to the low--energy YM theory.
%The pattern of supersymmetry breaking would be more
%fascinating if the perturbation 
It would be interesting to analyze other supersymmetry--breaking
deformations which
would not sit in the same direction of
the group space as the twisting of the normal bundle, which in
our case corresponds to the adjoint index $l = 3$ of $SU(2)_L$.
Such deformations would explicitly break the $U(1)$ R--symmetry, and
we showed that they 
%We showed that such a perturbation would necessarily
%break some of the symmetries of the background, or in other words 
would not be purely radial in seven dimensions, so they would be more
difficult to deal with. 

A different supergravity background, which is also related in some
limit to four--dimensional ${\cal N}=1$ pure SYM theory, was found in
\cite{Klebanov:2000hb}. It would be interesting to find supersymmetry
breaking deformations, similar to the one we discuss here, also for
that background. In that case the UV theory is less well--understood,
so it is not obvious \`a priori what supersymmetry--breaking deformations
can be performed.

\vspace{24pt}
{\large \bf Acknowledgments}

We would like to thank M. Berkooz and Y. Oz for useful discussions.
J.S. would like to thank J. Maldacena for a fruitful discussion.
O.A. would like to thank Tel-Aviv University for hospitality during
the course of this work.
This work was  supported in part by the US--Israel Binational Science
Foundation, by the Einstein Center at the Weizmann Institute, by Minerva, 
by the ISF Centers of Excellence program and by the European RTN network
HPRN-CT-2000-00122.
We would also like to thank the referee for bringing reference
\cite{Salam:1983fa} to our attention.

\newsec{Appendix A: Some $SO(4)$ Conventions}

We define a basis for the $SO(4)$ Lie algebra :
\begin{equation}
\alpha_L^1 = -i/2 \begin{pmatrix}  0 &  1 &  0 &  0 \\
                                  -1 &  0 &  0 &  0 \\
                                   0 &  0 &  0 &  1 \\
                                   0 &  0 & -1 &  0 
                  \end{pmatrix},\;\; 
\alpha_L^2 = -i/2 \begin{pmatrix}  0 &  0 &  1 &  0 \\
                                   0 &  0 &  0 & -1 \\
                                  -1 &  0 &  0 &  0 \\
                                   0 &  1 &  0 &  0 
                  \end{pmatrix},\;\;
\alpha_L^3 = -i/2 \begin{pmatrix}  0 &  0 &  0 &  1 \\
                                   0 &  0 &  1 &  0 \\
                                   0 & -1 &  0 &  0 \\
                                  -1 &  0 &  0 &  0 
                  \end{pmatrix},
\end{equation}
and
\begin{equation}
\alpha_R^1 = -i/2 \begin{pmatrix}  0 & -1 &  0 &  0 \\
                                   1 &  0 &  0 &  0 \\
                                   0 &  0 &  0 &  1 \\
                                   0 &  0 & -1 &  0 
                  \end{pmatrix},\;\; 
\alpha_R^2 = -i/2 \begin{pmatrix}  0 &  0 & -1 &  0 \\
                                   0 &  0 &  0 & -1 \\
                                   1 &  0 &  0 &  0 \\
                                   0 &  1 &  0 &  0 
                  \end{pmatrix},\;\;
\alpha_R^3 = -i/2 \begin{pmatrix}  0 &  0 &  0 & -1 \\
                                   0 &  0 &  1 &  0 \\
                                   0 & -1 &  0 &  0 \\
                                   1 &  0 &  0 &  0 
                  \end{pmatrix}.
\end{equation}
These conventions are closely related, but not identical, to those of
\cite{FreSch}. 
The commutators of those generators are
\begin{equation}
\left[ \alpha_L^l , \alpha_L^m \right] = i \epsilon_{lmn} \alpha_L^n
\end{equation}
and similarly for the $\alpha_R^r$, as needed for the $SU(2)$ algebra.
The anti--commutators are
\begin{equation}
\label{ALanticom}
\left\{ \alpha_L^l , \alpha_L^m \right\} = \half \delta_{lm} 
\end{equation}  
and similarly for the $\alpha_R^r$.

Now we define a basis for the $\bf(3,3)$ representation 
(acting on two $SO(4)$ vectors):
\begin{gather}
X_{11} = \begin{pmatrix} +1 &  0 &  0 &  0 \\
                          0 & +1 &  0 &  0 \\
                          0 &  0 & -1 &  0 \\
                          0 &  0 &  0 & -1 
         \end{pmatrix},\;\;
X_{12} = \begin{pmatrix}  0 &  0 &  0 & -1 \\
                          0 &  0 & +1 &  0 \\
                          0 & +1 &  0 &  0 \\
                         -1 &  0 &  0 &  0 
         \end{pmatrix},\;\;
X_{13} = \begin{pmatrix}  0 &  0 & +1 &  0 \\
                          0 &  0 &  0 & +1 \\
                         +1 &  0 &  0 &  0 \\
                          0 & +1 &  0 &  0 
         \end{pmatrix},\;\; \nonumber \\ \nonumber \\
X_{21} = \begin{pmatrix}  0 &  0 &  0 & +1 \\
                          0 &  0 & +1 &  0 \\
                          0 & +1 &  0 &  0 \\
                         +1 &  0 &  0 &  0 
         \end{pmatrix},\;\;
X_{22} = \begin{pmatrix} +1 &  0 &  0 &  0 \\
                          0 & -1 &  0 &  0 \\
                          0 &  0 & +1 &  0 \\
                          0 &  0 &  0 & -1 
         \end{pmatrix},\;\;
X_{23} = \begin{pmatrix}  0 & -1 &  0 &  0 \\
                         -1 &  0 &  0 &  0 \\
                          0 &  0 &  0 & +1 \\
                          0 &  0 & +1 &  0 
         \end{pmatrix},\;\; \nonumber \\ \nonumber \\
X_{31} = \begin{pmatrix}  0 &  0 & -1 &  0 \\
                          0 &  0 &  0 & +1 \\
                         -1 &  0 &  0 &  0 \\
                          0 & +1 &  0 &  0 
         \end{pmatrix},\;\;
X_{32} = \begin{pmatrix}  0 & +1 &  0 &  0 \\
                         +1 &  0 &  0 &  0 \\
                          0 &  0 &  0 & +1 \\
                          0 &  0 & +1 &  0 
         \end{pmatrix},\;\;
X_{33} = \begin{pmatrix} +1 &  0 &  0 &  0 \\
                          0 & -1 &  0 &  0 \\
                          0 &  0 & -1 &  0 \\
                          0 &  0 &  0 & +1 
         \end{pmatrix}.
\end{gather}
The matrix $X_{33}$, say, can be determined by demanding of the left
lowering operator $a_L^3 \equiv \alpha_L^1 - i \alpha_L^2$ that 
${a_L^3}^\dagger \, X_{33} \, a_L^3 = 0$, and similarly with 
$a_R^3$. The $X_{lr}$ can be seen to transform properly under
the rotations generated by $\alpha_L^l$ and $\alpha_R^r$. 
With those conventions, we have $X_{lr} = -4 \, \alpha_L^l \, \alpha_R^r$.

\newsec{Appendix B: Some Seven--Dimensional Supergravity Calculations}
In this appendix
we translate the various terms in the Lagrangian (\ref{L7preSO4})
into the $SU(2)_L \times SU(2)_R$ language. We work to quadratic
order in the fields $c_{lr}$, but we will comment on the way to get 
higher order expressions. 

In the kinetic term of the scalars in the ${\bf 9}$ representation,
in order to maintain only quadratic expressions in $c$, 
it is sufficient to substitute 
\begin{equation}
\tilde{T}_{ab} = \delta_{ab} + O(c)
\end{equation}
for $\tilde{T}$ (but not for its covariant derivative), since the derivatives
are at least of first order in $c$. Thus, this term takes the form
\begin{equation}
-\frac{1}{4} {\cal D}_\mu \tilde{T}_{ab} {\cal D}^\mu \tilde{T}_{ba} + O(c^3) =
-\frac{1}{4} \tr {\cal D}_\mu \tilde{T} {\cal D}^\mu \tilde{T} + O(c^3).
\end{equation} 
We now use the formula for the derivative of a matrix exponent, used
in the proof of the Baker--Campbell--Hausdorff formula,
\begin{equation}
\left( e^Q \right)' = e^Q (Q' - \frac{1}{2!} [Q , Q'] + 
                                \frac{1}{3!} [Q , [Q , Q']] - \cdots).
\end{equation}
For our purposes, it is sufficient to maintain simply
\begin{equation}
\tilde{T}' = Q' + O(c^2).
\end{equation}
Using (\ref{DT}) we get
\begin{equation}
{\cal D}_\mu \tilde{T} = \partial_\mu Q + \hg [A , Q] + O(c^2).
\end{equation}
Substituting we get that the kinetic term of the $c_{lr}$ fields is
\begin{equation}
- {\cal D}_\mu c_{lr} {\cal D}^\mu c_{lr},   
\end{equation}
where 
\begin{equation}
{\cal D}_\mu c_{lr} = \partial_\mu c_{lr} - 
  \sqrt{2} \hg (\epsilon_{lmn} A_L^m c_{nr} + \epsilon_{rst} A_R^s c_{lt}).
\end{equation}

For the Maxwell term, which can be written as 
\begin{equation}
\frac{1}{8} e^{-y/2} \tr F_{\mu \nu} \tilde{T}^{-1} F^{\mu \nu} 
\tilde{T}^{-1},
\end{equation}
one needs to substitute the quadratic
approximation of $\tilde{T}$,
\begin{equation}
\tilde{T}^{-1} = 1 - Q + \half Q^2 + O(c^3).
\end{equation} 
We find that this term is given by
\begin{equation}
-\frac{1}{4} e^{-y/2} (F_L^l F_L^l + F_R^r F_R^r + 4 F_L^l F_R^r c_{lr} 
         + 2 (F_L^l F_L^m c_{lr} c_{mr}  + F_R^r F_R^s c_{lr} c_{ls})) 
 + O(c^3),
\end{equation}
where, for example, $F_L^l F_L^l \equiv F_L^{l \mu \nu} F_{L \mu \nu}^l$.

Finally, the potential term is
\begin{equation}
V = -\half \hg^2 e^{y/2} (2 \tr \tilde{T}^2 - (\tr \tilde{T})^2).
\end{equation} 
Using the quadratic approximation of $\tilde{T}$ and (\ref{trQ})
one gets simply that
\begin{equation}
V = 4 \hg^2 e^{y/2} + O(c^3).
\end{equation}
If one were interested in higher orders of $c$ in the potential, one could
use the characteristic polynomial of $Q$ and the Vieta identities.
This polynomial,
$p(\lambda) = \det (\lambda 1 - Q)$, is an $SU(2)_L \times SU(2)_R$
invariant of $c_{lr}$, which is homogeneous of degree four in those
fields and in $\lambda$. Specifically,
\begin{equation}
\label{pQ}
p(\lambda) = \lambda^4 - 2 I_2 \lambda^2  - 8 I_3 \lambda + 2 I_4 - I_2^2,
\end{equation}
where the three invariants are
\begin{eqnarray}
I_2 & = & c_{lr} c_{lr} = \tr c c^T, \\
I_3 & = & \frac{1}{6} \epsilon_{lmn} \epsilon_{rst} c_{lr} c_{ms} c_{nt} = 
          \det c, \\
I_4 & = & c_{lr} c_{ls} c_{ms} c_{mr} = \tr c c^T c c^T.
\end{eqnarray}

\newsec{Appendix C: Some Six--Dimensional Supersymmetry Calculations}
We work with the following representation of the flat $SO(6)$ gamma
matrices:
\begin{gather}
\gamma^1 = i \begin{pmatrix}   
  0 &  0 &  0 &  0 &  0 &  0 &  0 & +1 \\ 
  0 &  0 &  0 &  0 &  0 &  0 & +1 &  0 \\ 
  0 &  0 &  0 &  0 &  0 & -1 &  0 &  0 \\ 
  0 &  0 &  0 &  0 & -1 &  0 &  0 &  0 \\ 
  0 &  0 &  0 & +1 &  0 &  0 &  0 &  0 \\ 
  0 &  0 & +1 &  0 &  0 &  0 &  0 &  0 \\ 
  0 & -1 &  0 &  0 &  0 &  0 &  0 &  0 \\ 
 -1 &  0 &  0 &  0 &  0 &  0 &  0 &  0   
             \end{pmatrix},\;\;
\gamma^2 = i \begin{pmatrix}
  0 &  0 &  0 & -1 &  0 &  0 &  0 &  0 \\ 
  0 &  0 & -1 &  0 &  0 &  0 &  0 &  0 \\ 
  0 & +1 &  0 &  0 &  0 &  0 &  0 &  0 \\ 
 +1 &  0 &  0 &  0 &  0 &  0 &  0 &  0 \\ 
  0 &  0 &  0 &  0 &  0 &  0 &  0 & +1 \\ 
  0 &  0 &  0 &  0 &  0 &  0 & +1 &  0 \\ 
  0 &  0 &  0 &  0 &  0 & -1 &  0 &  0 \\ 
  0 &  0 &  0 &  0 & -1 &  0 &  0 &  0   
             \end{pmatrix},\;\; \nonumber \\ \nonumber \\
\gamma^3 = i \begin{pmatrix}
  0 &  0 &  0 &  0 &  0 &  0 & -1 &  0 \\ 
  0 &  0 &  0 &  0 &  0 &  0 &  0 & +1 \\ 
  0 &  0 &  0 &  0 & -1 &  0 &  0 &  0 \\ 
  0 &  0 &  0 &  0 &  0 & +1 &  0 &  0 \\ 
  0 &  0 & +1 &  0 &  0 &  0 &  0 &  0 \\ 
  0 &  0 &  0 & -1 &  0 &  0 &  0 &  0 \\ 
 +1 &  0 &  0 &  0 &  0 &  0 &  0 &  0 \\ 
  0 & -1 &  0 &  0 &  0 &  0 &  0 &  0   
             \end{pmatrix},\;\;
\gamma^4 = i \begin{pmatrix}
  0 &  0 & -1 &  0 &  0 &  0 &  0 &  0 \\ 
  0 &  0 &  0 & +1 &  0 &  0 &  0 &  0 \\ 
 +1 &  0 &  0 &  0 &  0 &  0 &  0 &  0 \\ 
  0 & -1 &  0 &  0 &  0 &  0 &  0 &  0 \\ 
  0 &  0 &  0 &  0 &  0 &  0 & -1 &  0 \\ 
  0 &  0 &  0 &  0 &  0 &  0 &  0 & +1 \\ 
  0 &  0 &  0 &  0 & +1 &  0 &  0 &  0 \\ 
  0 &  0 &  0 &  0 &  0 & -1 &  0 &  0   
             \end{pmatrix},\;\; \nonumber \\ \nonumber \\
\gamma^5 = i \begin{pmatrix}
  0 &  0 &  0 &  0 &  0 & +1 &  0 &  0 \\ 
  0 &  0 &  0 &  0 & +1 &  0 &  0 &  0 \\ 
  0 &  0 &  0 &  0 &  0 &  0 &  0 & +1 \\ 
  0 &  0 &  0 &  0 &  0 &  0 & +1 &  0 \\ 
  0 & -1 &  0 &  0 &  0 &  0 &  0 &  0 \\ 
 -1 &  0 &  0 &  0 &  0 &  0 &  0 &  0 \\ 
  0 &  0 &  0 & -1 &  0 &  0 &  0 &  0 \\ 
  0 &  0 & -1 &  0 &  0 &  0 &  0 &  0   
             \end{pmatrix},\;\;
\gamma^6 = i \begin{pmatrix}
  0 & +1 &  0 &  0 &  0 &  0 &  0 &  0 \\ 
 -1 &  0 &  0 &  0 &  0 &  0 &  0 &  0 \\ 
  0 &  0 &  0 & +1 &  0 &  0 &  0 &  0 \\ 
  0 &  0 & -1 &  0 &  0 &  0 &  0 &  0 \\ 
  0 &  0 &  0 &  0 &  0 & +1 &  0 &  0 \\ 
  0 &  0 &  0 &  0 & -1 &  0 &  0 &  0 \\ 
  0 &  0 &  0 &  0 &  0 &  0 &  0 & +1 \\  
  0 &  0 &  0 &  0 &  0 &  0 & -1 &  0   
             \end{pmatrix}.
\end{gather}
Then, $M^{(0)}$ defined in section 8 is given by
\begin{equation}
M^{(0)} = P \, 
  \begin{pmatrix}
       0 & 0 & 0 & 0 & 1 & 0 & 4 \rho & -1 + 4 \rho \\ 
       0 & 0 & 0 & 0 & 0 & 1 & -1 + 4 \rho & 4 \rho \\ 
       0 & 0 & 0 & 0 & -4 \rho & 1 - 4 \rho & -1 & 0 \\ 
       0 & 0 & 0 & 0 & 1 - 4 \rho & -4 \rho & 0 & -1 \\ 
       1 & 0 & -4 \rho & -1 + 4 \rho & 0 & 0 & 0 & 0 \\ 
       0 & 1 & -1 + 4 \rho & -4 \rho & 0 & 0 & 0 & 0 \\ 
       4 \rho & 1 - 4 \rho & -1 & 0 & 0 & 0 & 0 & 0 \\ 
       1 - 4 \rho & 4 \rho & 0 & -1 & 0 & 0 & 0 & 0
  \end{pmatrix},
\end{equation}
where the prefactor is
\begin{equation}
P = \frac{1}{8 \sqrt{2}} \, g \, e^{\varphi/4} \, \rho^{-1},
\end{equation}
and for $\varphi$ we should take the unperturbed value given by
(\ref{singphi}). 

The kernel of $M^{(0)}$ is given by 
$\Ker M^{(0)} = \Span \{\zeta_1 , \zeta_2\}$
where we may take
\begin{eqnarray}
\zeta^1 & = & \left( +1, +1, +1, +1,  0,  0,  0,  0 \right)^T, \\
\zeta^2 & = & \left(  0,  0,  0,  0, +1, -1, -1, +1 \right)^T. 
\end{eqnarray}
The orthogonal space to the image of $M^{(0)}$ is given by
$\left(\Image M^{(0)}\right)^\bot = \Span \{\xi_1 , \xi_2\}$ 
where we may take
\begin{eqnarray}
\xi^1   & = & \left(  0,  0,  0,  0, -1, -1, +1, +1 \right), \\
\xi^2   & = & \left( -1, +1, -1, +1,  0,  0,  0,  0 \right). 
\end{eqnarray}

The matrix $M^{(1)}$ is too cumbersome to write explicitly, but we have
\begin{eqnarray}
\xi^1 M^{(1)} \zeta^1 = \xi^2 M^{(1)} \zeta^2   & = & 
       -4 \cos \ttheta \, P \, 
          \left( 2 \rho \, c_3'(\rho) + (1 + 4 \rho) \, c_3(\rho) \right), \\
\xi^2 M^{(1)} \zeta^1 = \xi^1 M^{(1)} \zeta^2   & = & 0,
\end{eqnarray}
so the necessary condition for supersymmetry to be conserved is
given by (\ref{susycondition}).

%%%%%%%%%%%%%%%%%%%%%%%%%%%%%%%%%%%%%%%%%%%%%%%%%%%%%%%%%%%%%%%%%%%%%%%%%%%%%

\end{document}